\documentclass{aa}

\usepackage{txfonts} \usepackage{color}

\usepackage{graphicx} \usepackage{amssymb}
\usepackage{array} \usepackage{wrapfig} \usepackage{subfig}
\usepackage{placeins} \usepackage{url}

\usepackage{natbib} \usepackage{cite}
\bibpunct{(}{)}{;}{a}{}{,}

\def\aap{{A\&A}}                

\title{Long-term optical variability of PKS~2155-304}
\author{M.~A.~ Kastendieck\inst{\ref{inst1}} \and M.~C.~B.~
Ashley\inst{\ref{inst2}} \and D.~ Horns\inst{\ref{inst3}}}
\institute{ Institute for Experimental Physics, Univ. of
Hamburg, Luruper Chaussee 149, D-22761 Hamburg, Germany\\
\email{max.kastendieck@physik.uni-hamburg.de}\label{inst1}
\and School of Physics, Univ. of New South Wales, NSW 2052,
Australia\\ \email{m.ashley@unsw.edu.au}\label{inst2} \and
Institute for Experimental Physics, Univ. of Hamburg,
Luruper Chaussee 149, D-22761 Hamburg, Germany\\
\email{dieter.horns@physik.uni-hamburg.de}\label{inst3} }
\date{Received October 12, 2010 / Accepted May 09, 2011}
\abstract {
} {
The optical variability of the blazar PKS~2155-304 is
investigated to characterise the red noise behaviour at
largely different time scales from 20 days to ${\cal
O}(>10\mathrm{yrs})$.  } {
The long-term optical light curve of PKS~2155-304 is
assembled from archival data as well as from so-far
unpublished observations mostly carried out with the
ROTSE-III and the ASAS robotic telescopes.  A forward
folding technique is used to determine the best-fit
parameters for a model of a power law with a break in the
power spectral density function (PSD). The best-fit
parameters are estimated using a maximum-likelihood method
with simulated light curves in conjunction with the Lomb
Scargle Periodogram and the first-order Structure Function
(SF). In addition, a new approach based upon the so-called
Multiple Fragments Variance Function (MFVF) is introduced
and compared to the other methods. Simulated light curves
have been used to confirm the reliability of these methods
as well as to estimate the uncertainties of the best-fit
parameters.  } {
The light curve is consistent with the assumed broken
power-law PSD. All three methods agree within the estimated
uncertainties with the MFVF providing the most accurate
results. The red-noise behaviour of the PSD in frequency $f$
follows a power law with $f^{-\beta}$,
$\beta=1.8^{+0.1}_{-0.2}$ and a break towards $f^{0}$ at
frequencies lower than
$f_\mathrm{min}=(2.7^{+2.2}_{-1.6}~\mathrm{yrs})^{-1}$.  } {
} 

\begin{document} \maketitle \section{Introduction} It is
widely accepted that the broad-band emission of an active
nucleus of a galaxy (AGN)  is largely dominated by
non-thermal (synchrotron and inverse Compton) emission from
relativistic electrons or positrons (or protons)
\citep{1997ARA&A..35..445U}.  In the optical band, thermal
emission from the central region (broad or narrow-line
regions, accretion disk) and stellar emission from the host
galaxy additionally contribute.  In blazars (to which
PKS~2155-304 belongs) the jet emission is commonly
considered to be dominant over all other (thermal)
contributions consistent with the absence of prominent
emission or absorption lines in the optical spectrum.  While
it is still an open question how and if changes in the
accretion rate propagate into variations of the jet
emission, there are two common explanations for the
variability to be observed from the jet: self-organised
criticality (SOC) \citep{1987PhRvL..59..381B}, either from
the accretion process or jet process resulting in a red
noise behaviour, and jet dynamics causing periodic
oscillations \citep{2004ApJ...615L...5R}.

 Over sufficiently long time scales the average flow of
matter drifting from the outer region of the accretion disk
towards the inner most stable orbit is believed to be
constant. On smaller time scales, the matter may fall in
avalanches of different size and duration towards the centre
\citep{1999PhST...82..133D}.  \citet{1999IAUS..194..356K}
showed that this can be simulated in a simple automaton
model producing SOC. If the optical emission correlates to
the released energy of the avalanches, this model results in
a red noise light curve.  A more general motivation on SOC
in BL Lac objects is given by the models of SOC in compact
plasmas \citep{1998ApJ...503L..57S} and astronomical shocks
\citep{2000ApJ...533L.171M}.  At long time scales the effect
of SOC may be limited by the effective size of the emitting
region.  Therefore, it is expected that at longer time
scales the SOC and therefore the red noise behaviour
vanishes.

\object{PKS 2155-304} is an optical bright X-ray selected BL
Lac type object ($m_\mathrm{V} \approx 14^\mathrm{mag}$) at
a redshift of $z=0.116$ \citep{1993ApJ...411L..63F}.  Ground
based observations in the near-infrared have been used to
separate the relatively faint
($M_\mathrm{r}=-24.4^\mathrm{mag}$) elliptical host galaxy
from the bright nucleus \citep{1998A&A...336..479K}.
\citet{1992ApJ...400..115S} and \citet{1997A&A...325...27H}
observed a variation of the optical flux of $2\,\%$ and $10
\, \%$ respectively per day.  \citet{1997A&A...327..539P}
found fast variations on the time scales of $t_\mathrm{min}
< 15 \,\textrm{min}$ which were limited by the noise of the
instruments.  They also found that on a time scale of days
the variability is compatible with red noise with a
power-law index of $\beta=2.4$.  X-ray observations indicate
a similar power-law behaviour ranging from $1.7$ to $3.5$
during several continuous pointings lasting up to a day
\citep{2010ApJ...718..279G}.  Recently, red noise has been
detected from PKS~2155-304 at very high energies
($>100$~GeV) on timescales of hours up to frequencies of
$\frac{1}{600}\,\textrm{Hz}$ in the power spectral density
function \citep{2007ApJ...664L..71A}.

Searches for periodicities in the optical revealed evidences
of a periodicity of $0.7 \, \textrm{days}$
\citep{1993ApJ...411..614U}.

In this paper a detailed $\approx7$~years long light curve
of PKS~2155-304 measured with the ROTSE-III telescope system
is presented. Archival and almost simultaneous observations
are added to the data to obtain a long-term light curve
covering timescales up to $\approx 77$~years. Subsequently
the intrinsic power spectral density function (PSD) is
analysed with the Lomb Scargle Periodogram, the Structure
Function and a new method called Multiple Fragments Variance
Function. The result is compared to the results of simulated
light curves with red noise behaviour with different shapes
of the PSDs. The best-fit parameter set for the assumed
model is found with a maximum-likelihood analysis. 

In Section~\ref{sect:observations}, a summary of the
observations is given.  In
Section~\ref{sect:time_series_analysis} the methods and
maximum-likelihood analysis used to characterise the
long-term behaviour of the optical flux of PKS~2155-304 are
described in detail.  Readers who are mainly interested in
the results, may go directly to Section~\ref{sec:results},
where the long-term light curve of the object and the
results of the methods and maximum-likelihood analysis are
presented. The paper closes with a conclusion in
Section~\ref{sec:summary}.

\section{Summary of observations} \label{sect:observations}
\subsection{The ROTSE telescope system and data analysis}
The ROTSE-III telescope
system\footnote{\url{http://www.rotse.net/}} consists of
four robotic telescopes (called ROTSE-IIIa, b, c and d)
located in Australia, USA, Namibia, and Turkey.  The
telescopes are of a Ritchey-Chr\'etien type with a parabolic
primary mirror ($\diameter\, 450 \,\textrm{mm}$) and a focal
length of $850 \,\textrm{mm}$.  The field of view
encompasses $1.85^\circ \times 1.85^\circ$.  Only light with
a wavelength between $400$ and $900 \,\textrm{nm}$ can pass
the broad-band filter.  On average each telescope collects
$250$ observational frames per night.  Typically, the
limiting magnitude is $\approx 17.3^\mathrm{mag}$ (up to
$19^\mathrm{mag}$ at perfect conditions) and a few thousand
individual objects are detected within each frame. ROTSE-III
is equipped with a fully automated computer-controlled data
acquisition system (DAQ) as well as an automatic
data-reduction pipeline.  \citet{2003PASP..115..132A}
provides more details on the system.

The observations used for this analysis have been carried
out whenever PKS~2155-304 was observable. Observations are
taken as pairs of exposures of 20 or 60~s each with a
dithering pointing pattern to reduce the impact of hot
pixels. Pairs of observations are taken at latency intervals
of approximately 45 minutes.

Each frame is dark-field subtracted and flat-field
corrected.  The thin entrance window makes an additional
fringe-pattern correction necessary.  After the objects in
the frame have been extracted by means of the
\textit{SExtractor} \citep{sextractor_anleitung} software,
their coordinates and instrumental (uncalibrated) magnitudes
are matched and adjusted to the USNO
A2.0\footnote{\url{http://tdc-www.harvard.edu/catalogs/ua2.html}}
catalogue.  During this process, individual objects are
flagged as problematic (e.g.  blended point-spread
functions, saturated pixels etc.).  \\ The photometric data
of all objects detected in a sequence of frames taken on a
field are combined in so-called match-structures.  During
this process, a relative correction of fluctuations in the
photometry is applied, which benefits from the large field
of view with a correspondingly larger number of detected
objects.  Details on the data calibration and analyses chain
are provided in \citet{rotse_anleitung}.

\begin{table} \caption{Data log of the ROTSE observations.}
\label{tab_datalog} \centering \begin{tabular}{l c} \hline
\hline Total number of frames & 12\,946 \\ Number of frames
with sufficient quality & 6\,310 \\ \hline Average flux in
mJy & 15.9 \\ Standard deviation in mJy & 6.9\\ \hline
\end{tabular} \end{table} 

Starting August 24th, 2003, PKS~2155-304 has been observed
regularly with the ROTSE-IIIc telescope system in Namibia.
Additional observations have been performed with ROTSE-IIIa
in Australia since October 10th, 2003. Data up to October
26th, 2010 are presented and analysed in this work.  Table
\ref{tab_datalog} shows statistics of the data on
PKS~2155-304.  To assure high data quality and reduce
systematic errors, images with inferior quality reported by
the DAQ are rejected.  The complete light curve is presented
in Section \ref{sec:results_lightcurve}, Fig.
\ref{fig_rotse_lightcurve}. A table containing the data is
available at the CDS. Column 1 lists the Julian date of the
observation, Column 2 and 3 give the apparent luminosity
(R-Band) with the statistical uncertainty.
 
Since ROTSE does not take use of an R-band filter but
calibrates the instrumental magnitude to the catalogue
values of stars, the measured R-magnitudes of PKS~2155-304
suffer from a systematic deviation due to the non-star like
spectrum.  This requires a cross calibration with
simultaneous R-band observations taken with other
instruments \citep[e.g. the ATOM telescope
in][]{2009ApJ...696L.150A}.  

\subsection{Archival data and observations} \begin{table}
\caption{Overview of the archival data used in this work.}
\label{tab_archival_data} \centering \begin{tabular}{c c c}
\hline \hline Period & Number of data points & Reference \\
\hline 1934 - 1940 & 61 & $(1)$ \\ 1979 - 1995 & 134 &
$(2)\dots(13)$ \\ 1980
- 2007 & 1\,721 & $(14)\dots(16)$ \\ 1998 - 1999 & 12 &
  $(17)$ \\ \hline \end{tabular} \tablebib{ (1)~
\citet{1979ApJ...234..810G};
(2)~\citet{1979ApJ...234..810G},
(3)~\citet{1983ApJ...272...26M},
(4)~\citet{1986MNRAS.221..739B},
(5)~\citet{1987A&AS...68..383H},
(6)~\citet{1988AJ.....96.1215P},
(7)~\citet{1989ApJ...341..733T},
(8)~\citet{1990A&AS...83..183M},
(9)~\citet{1991ApJ...383..580S},
(10)~\citet{1992ApJ...385..146C},
(11)~\citet{1992ApJ...400..115S},
(12)~\citet{1993ApJS...85..265J} and
(13)~\citet{1993ApJ...411..614U} as collected in
\citet{1996A&AS..116..289Z};
(14)~\citet{2004PhDT........28C},
(15)~\citet{1995ApJ...438..108C} and
(16)~\citet{1997ApJ...486..770P} as collected in
\citet{2007ApJ...671...97O};
(17)~\citet{2001ApJS..132...73T}.  } \end{table} Archival
(mostly photometric plate) data on PKS~2155-304 covering the
period from 1934 to 2007 have been collected from several
publications as listed in Table \ref{tab_archival_data}.

Furthermore,  $397$ mostly simultaneous observations have
been obtained with the \textit{All Sky Automated Survey}
(ASAS) \citep{2002AcA....52..397P} from November 25th, 2000
till October 28th, 2008.  ASAS is a project dedicated to
continuous photometric monitoring of  approximately $10^7$
stars brighter than 14 magnitude to search for transient and
variable objects.  The data used here were taken with the
ASAS-2 instrument located at Las Campanas, Chile.  The
ASAS-2 consists of a $f=135$~mm, $f/D=1/1.8$ lens system and
a Pictor 416 with a \textit{KODAK} KAF-0400 CCD sensor
(512x768 pixel). ASAS has been monitoring the southern
hemisphere since 2000. 

Where required the magnitudes given for a different
wavelength band are adjusted to the R-band by adding a
constant offset obtained from observations taken with the
ATOM telescope \citep{2009ApJ...696L.150A}. The apparent
luminosities of the whole data set (ROTSE, ASAS and archival
data) are converted into flux units with $13.3^\mathrm{mag}$
corresponding to $10$~mJy \citep{2005A&A...442..895A}.

\subsection{Statistical and systematic uncertainties of the
photometry} \label{sec:uncertainties_of_obs} 
\begin{figure}[htb!]
\includegraphics[angle=90,width=\linewidth]{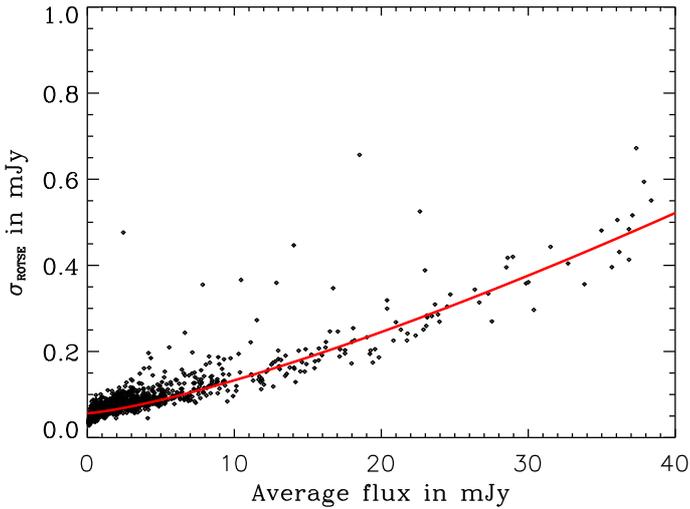}
\caption{Standard deviations $\sigma_\mathrm{ROTSE}$ of the
light curves of $\approx 3\,000$ individual objects
represented by the data points.  The red curve is the
least-squares fit to the medians of the values within
$1$~mJy bins.} \label{fig_errors_flux} \end{figure}

The estimated
statistical error is crucial for simulating the detector
noise behaviour, and therefore we have carried out a careful
analysis of the detector noise as a function of the observed
brightness.  For the ROTSE-III data, we have estimated the
total (statistical and systematic error) from the data:
based upon $660$ individual pointings on PKS~2155-304, we
have selected $\approx 3\,000$ other objects that are
covered in these pointings.  For each of these $3\,000$
light curves, the square root of the variance
$\sigma_\mathrm{ROTSE}$ has been calculated. The correlation
between $\sigma_{ROTSE}$ and the flux $\Phi$ of an object
follows a power-law shape which has been parametrised with
\begin{eqnarray} \label{eqn:rotse_noise}
\frac{\sigma_\mathrm{ROTSE}(\Phi)}{\mathrm{mJy}} &=&
0.0764\left(\frac{\Phi}{10\,\mathrm{mJy}}\right)^{1.302}+0.0564.
\end{eqnarray} In order to neglect any variable objects
median values have been used to perform a least-squares fit.
The median values are calculated from the values within bins
of $1$~mJy. The values and the fitted power-law are shown in
Fig.\ref{fig_errors_flux}

For the ASAS data,  the quoted photometric uncertainty has
been parametrised as a function of the total flux:

 \begin{eqnarray}
\frac{\sigma_\mathrm{ASAS}(\Phi)}{\mathrm{mJy}} &=&
0.162\left(\frac{\Phi}{10\,\mathrm{mJy}}\right)^{1.128}.
\end{eqnarray}

The photometric uncertainty on the archival plate data has
been quoted to be in the range of $0.01\ldots 0.1$
magnitudes.  These values are used to simulate the absolute
uncertainties for the corresponding simulated data points.

\section{Time series analysis}
\label{sect:time_series_analysis} Two well-established
methods -- the Lomb-Scargle Periodogram (LSP) (see Appendix
\ref{sub:LSP}) and the Structure Function (SF) (see Appendix
\ref{sub:SF}) -- as well as a newly developed method called
Multiple Fragments Variance Function (MFVF) (see Section
\ref{sub:mlvf}) are used to characterise the timing
behaviour of the optical long-term light curve of
PKS~2155-304.

\subsection{Multiple Fragments Variance Function (MFVF)}
\label{sub:mlvf} \begin{figure}[htb!]
\includegraphics[width=\linewidth]{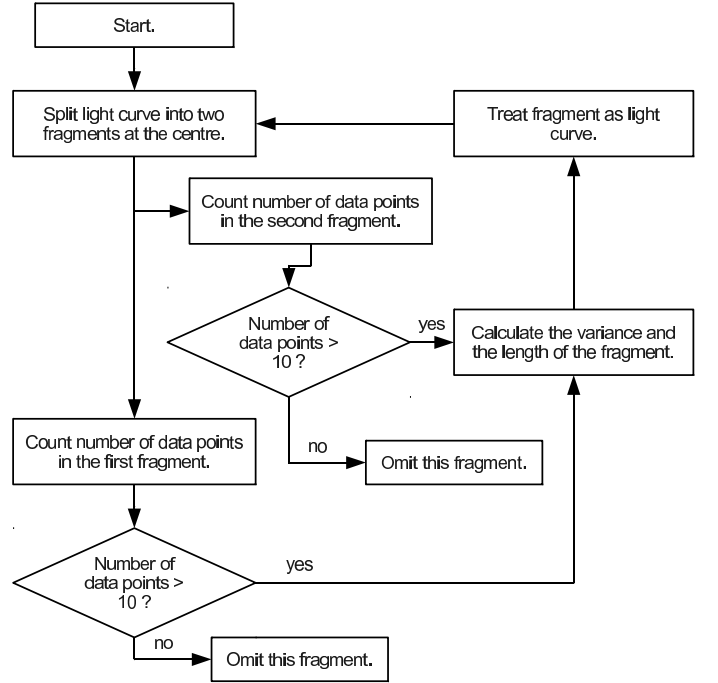}
\caption{Flowchart of the Multiple Fragments Variance
Function (MFVF).} \label{fig_mfv_diagram} \end{figure} We
have developed an alternative method to characterise the
variability of the source at different timescales in a
simple and robust way. The variance \begin{eqnarray}
\mathrm{var}(a)&:=&\langle (\langle a\rangle-a)^2\rangle
\end{eqnarray} of the entire light curve is computed and
subsequently, the light curve is subdivided at the half of
the time lengths of the interval into two fragments and the
variance for each of the two independent fragments is
computed\footnote{Alternatively the interval may be split at
the position of the largest gap.}.  This process of
subdividing the light curve continues recursively until the
number of remaining measurements within the fragment drops
below a critical value (see Fig.  \ref{fig_mfv_diagram} for
a flowchart of this process).  For the analysis carried out
here, the requirement on the minimum number of measurements
within the fragment is chosen to be $> 10$. This number is a
compromise between the limited number of measurements and
the smallest timescale resolvable.

The fragment length is calculated by the time difference
between the first and last measurement available within this
fragment.  The variance is therefore calculated for light
curve segments which cover intervals with widely different
values of duration.  Finally, these values are averaged
(binned) within 100 bins per decade of fragment length. This
function yields information about the variance of the
variability depending on the time scale. 

Like the SF the MFVF is computed in the time domain.  Unlike
the other methods it is not sensitive to periodicities. It
rather concentrates on information about the mean
variability amplitude of the light curve occurring on a
specific time scale.  Therefore it is a goal-oriented method
to find a maximum time scale of the variability.  The
sampling of the light curve is taken directly into account,
so the distorting effect of uneven sampling is limited to an
acceptable level (see below). An IDL as well as a Python
script for the MFVF can be found at
\url{https://sourceforge.net/projects/mfvf/}.

\subsection{Comparison of the methods with simulated light
curves} \label{sec:comparison_sim_lc} Since all methods
suffer at various degrees from the effect of unevenly
sampled light-curves (sampling effects), we use sets of
simulated light-curves, in order to take sampling effects
and the influence of instrumental noise into account. These
simulated light-curves are useful to compare the behaviour
of the three methods.  

The simulated light curves are generated under the
assumption of a red noise with a power spectral density
function (PSD) following a power-law ($\mathcal{PSD}\propto
f^{-\beta}$ above a characteristic frequency
$f_\mathrm{min}$).  The original recipe for generating pure
red-noise light-curves first introduced by
\citet{1995A&A...300..707T} has been extended to include a
break at $f_\mathrm{min}$. For frequencies below
$f_\mathrm{min}$, we assume a white noise behaviour
($\mathcal{PSD}=const.$).  The normalisation of the
simulated light curve is chosen to match the mean flux level
and the variance of the measured light curve.  \\ In order
to investigate sampling effects the light curves were
generated either with an even sampling (ideal case) or with
an uneven sampling (realistic case).  For both cases the
total length $\Delta t_\mathrm{total}$ of the light curve
and the number of data points $N_\mathrm{obs}$ are chosen to
match the values of the light curve of PKS~2155-304. For the
ideal case the observation times are spread evenly over
$\Delta t_\mathrm{total}$.  This results in an evenly
sampled light curve with one measurement every $\Delta
t_\mathrm{total}/N_\mathrm{obs}=3.37$~days and without any
gaps.  

For the realistic case the sampling is chosen to be exactly
the same as it is for the light curve of PKS~2155-304.  The
raw evenly sampled simulated light curves are sampled down
to the actual measurement times.  Only for the realistic
case each individual flux measurement has been randomised
with a Gaussian function in order to account for
experimental uncertainties on the measurements (see Section
\ref{sec:uncertainties_of_obs} for details on these
uncertainties). For each data point the variance of the
Gaussian errors were chosen individually to simulate the
according instrumental noise.

For each case (ideal and realistic), 5\,000 independent
light curves (covering $\approx 27\,000$ days each) have
been simulated with $\beta=2.0$ and
$f_\mathrm{min}=10^{-3}$~d$^{-1}$.  For the ideal case, all
methods should produce results which are close to the
intrinsic behaviour of the light curve.  Here, the sampling
effects are limited to the periodic sampling and to the
finite length of the simulated light curves.  By comparing
the results from the realistic and ideal case simulations,
sampling effects related to the uneven sampling can be
investigated. 

\begin{figure*}[htb!]
\centerline{\includegraphics[angle=90,width=0.5\linewidth]{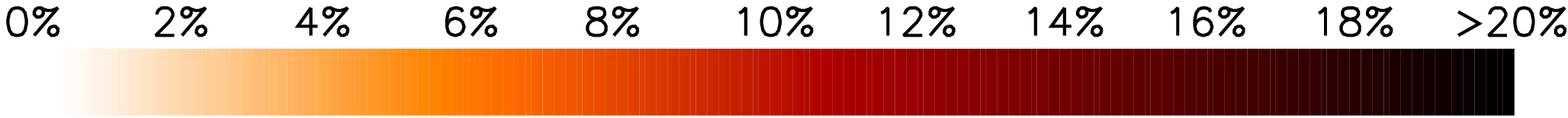}}
\vskip 6pt
\includegraphics[angle=90,width=0.5\linewidth]{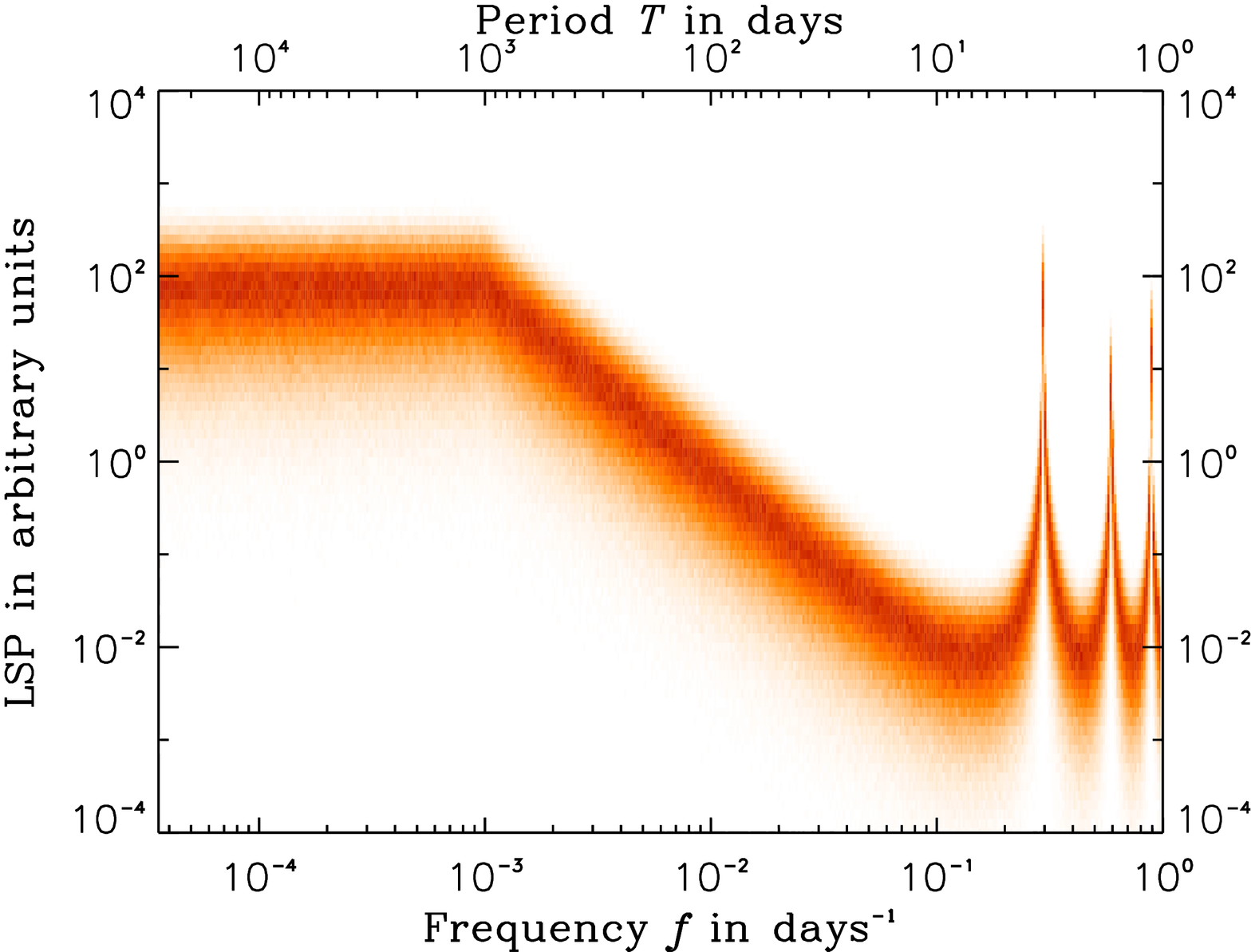}
\includegraphics[angle=90,width=0.5\linewidth]{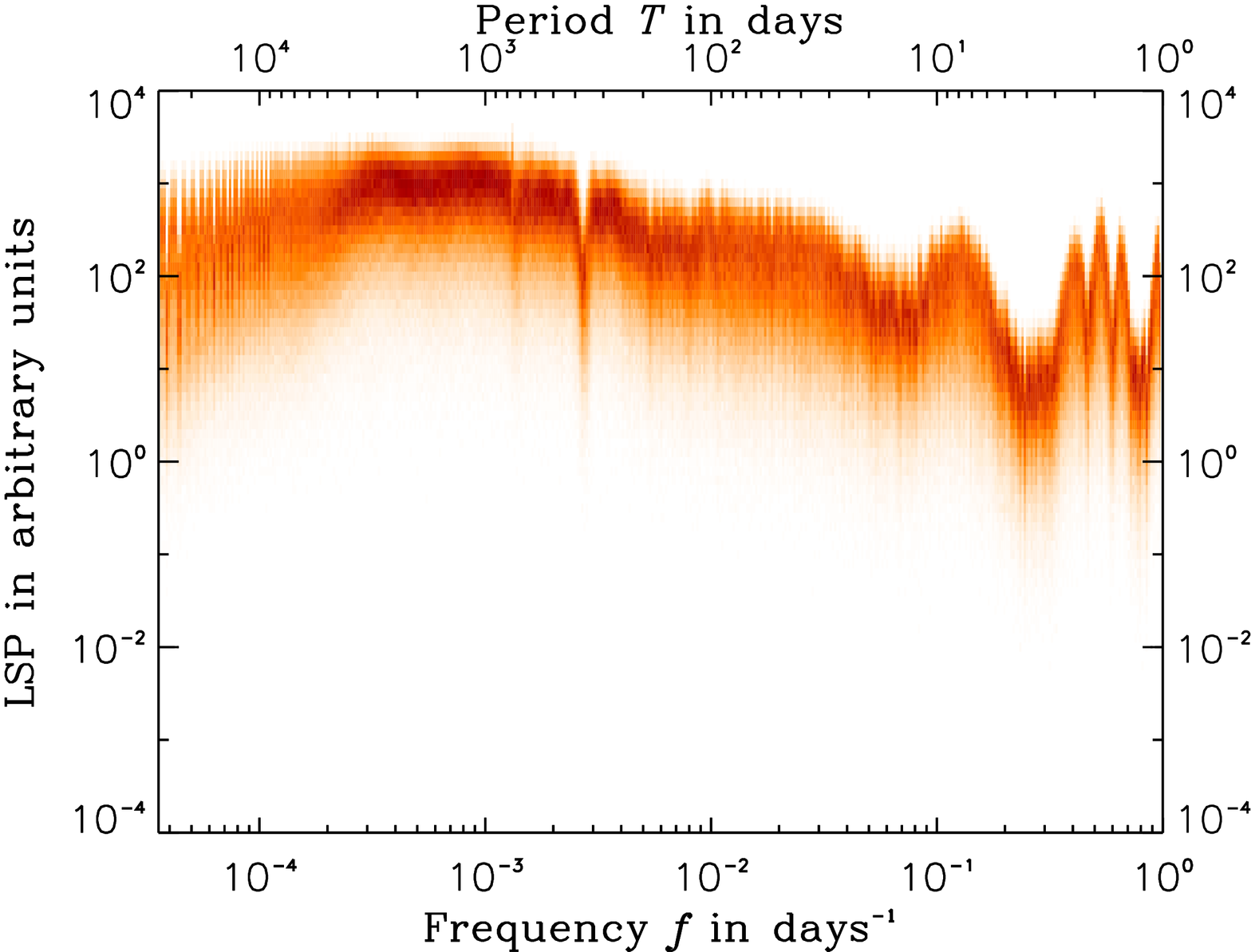}
\vskip 12pt
\includegraphics[angle=90,width=0.5\linewidth]{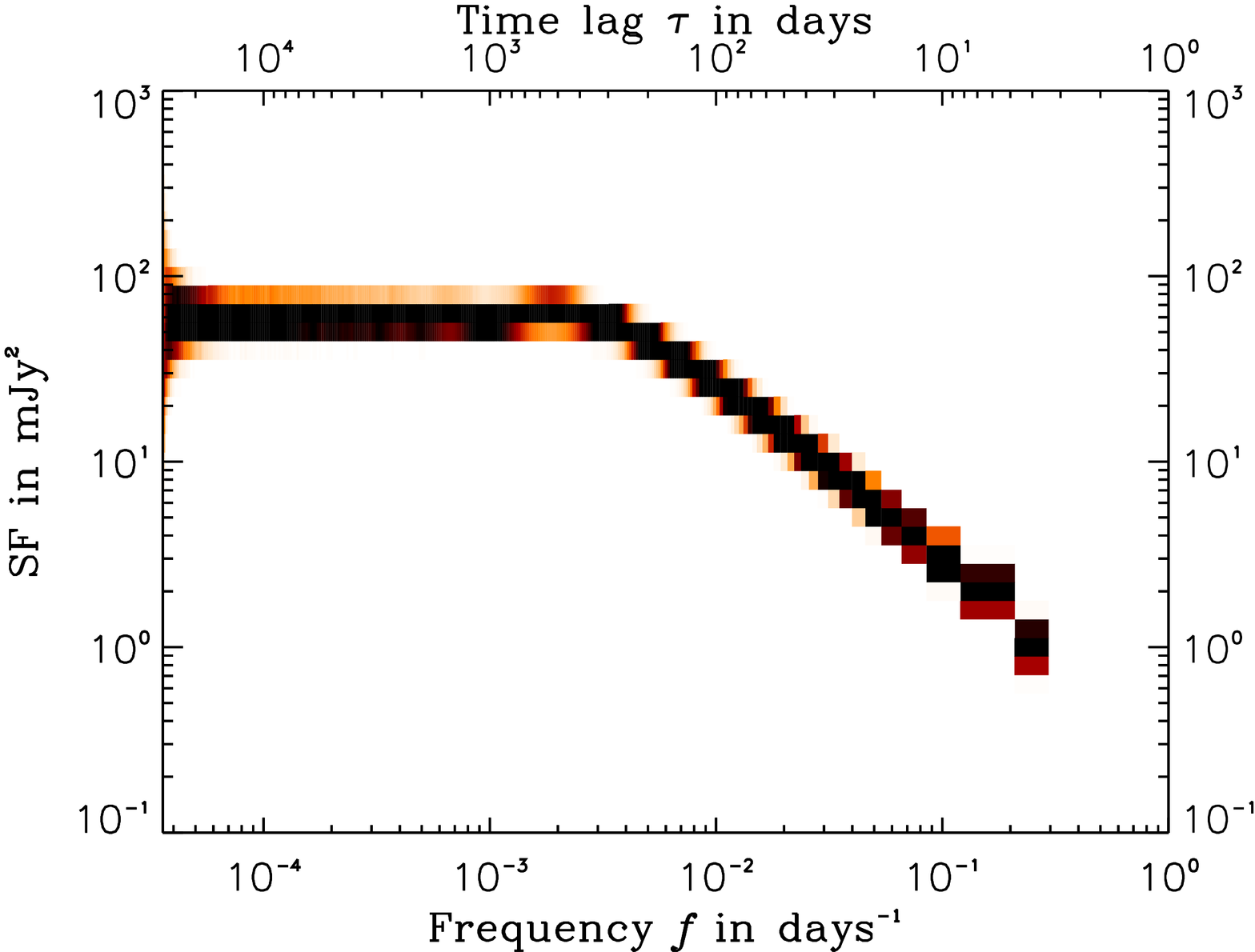}
\includegraphics[angle=90,width=0.5\linewidth]{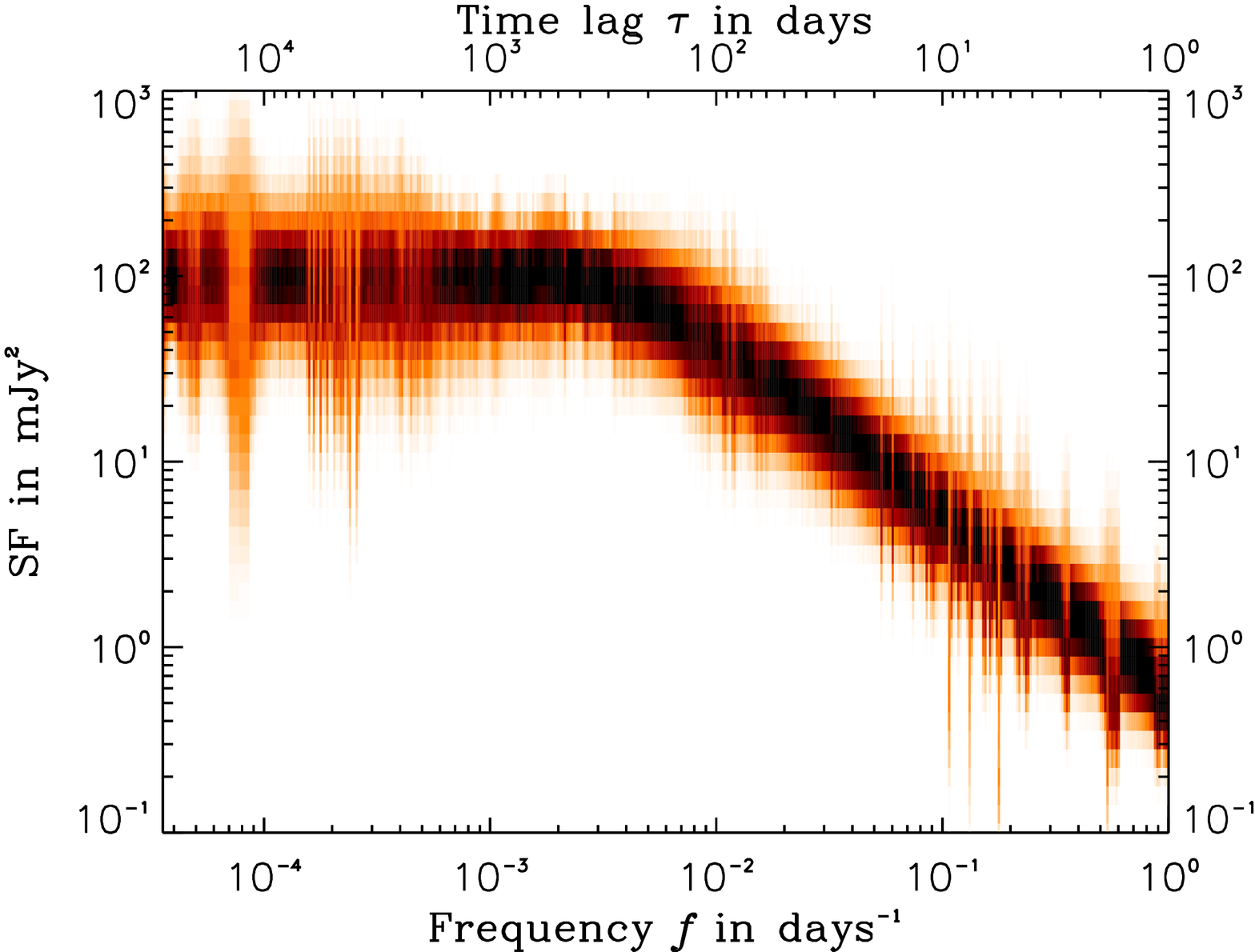}
\vskip 12pt
\includegraphics[angle=90,width=0.5\linewidth]{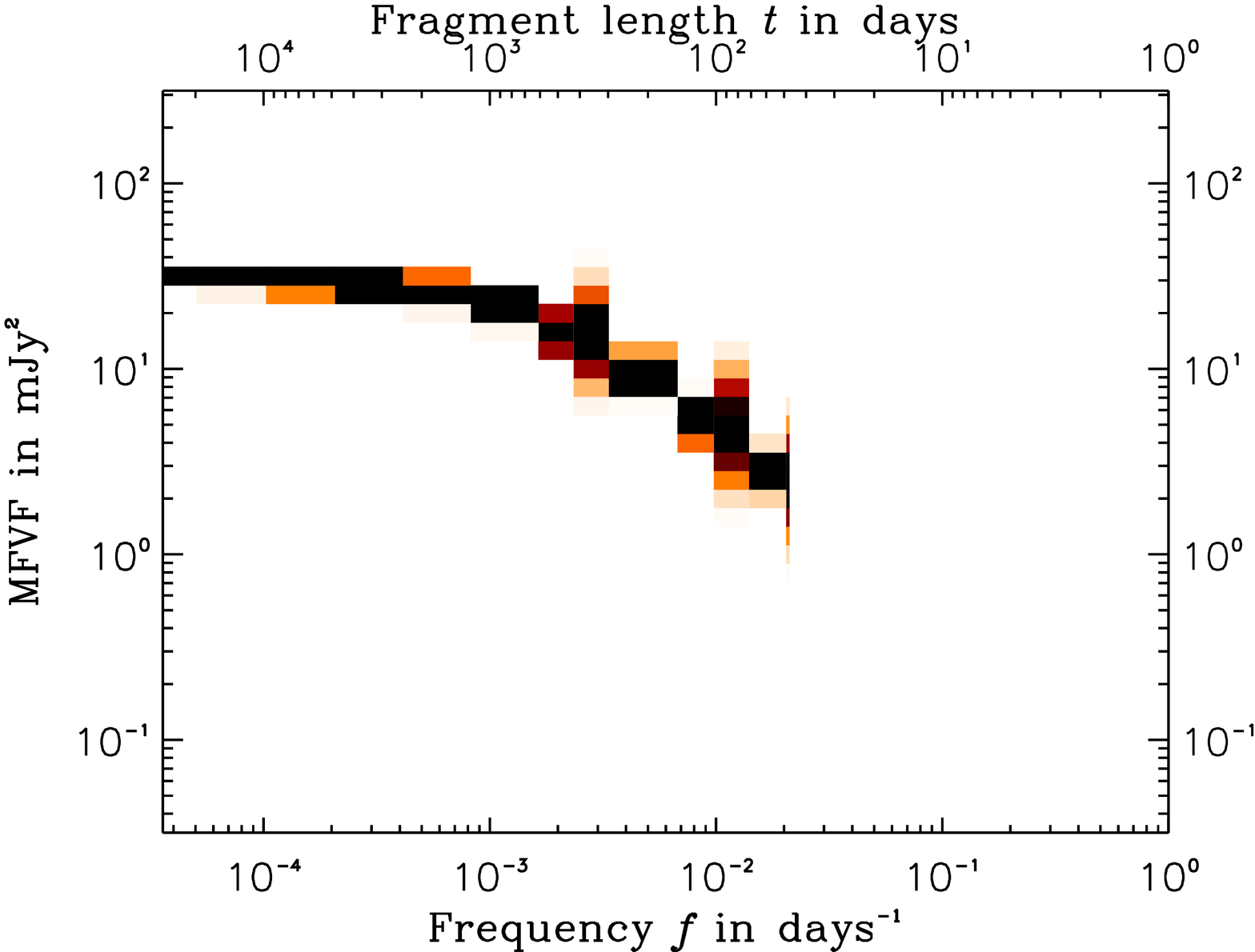}
\includegraphics[angle=90,width=0.5\linewidth]{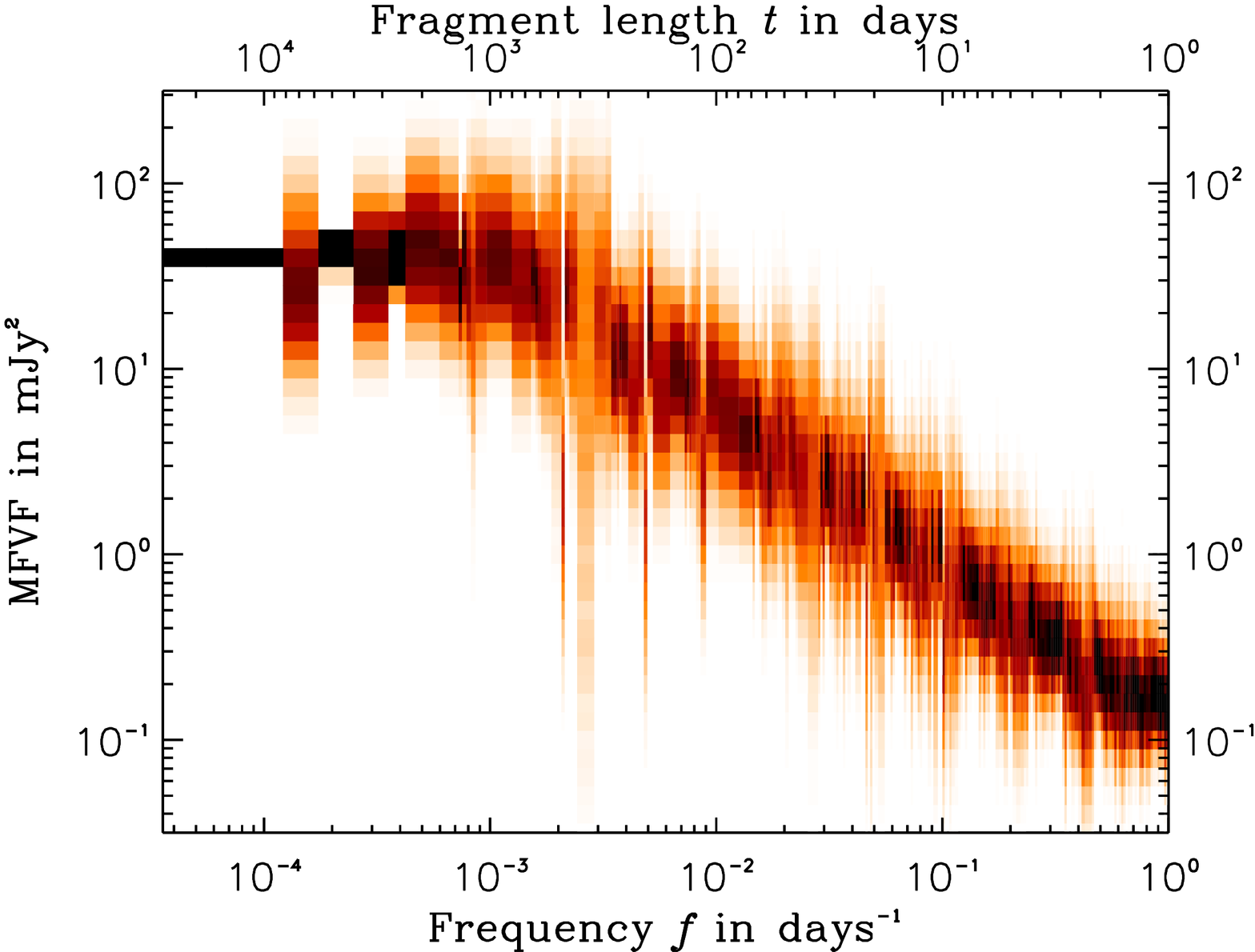}
\caption{Two dimensional histograms of LSP, SF, and MFVF
(from top to bottom) values of artificial light curves with
$\beta=2.0$ and $f_\mathrm{min}=10^{-3}$~d$^{-1}$. On the
left side histograms are shown for the case of an even
sampling of one measurement every 3.37~days while on the
right side, the effects of a realistic (uneven) sampling and
simulated instrumental noise are visible (see text for
details). } \label{fig_2dhistogram} \end{figure*}

The individual values of the methods are accumulated in
two-dimensional histograms.  In Fig.~\ref{fig_2dhistogram},
the histograms for the LSP (upper panel), SF (middle panel),
and MFVF (lower panel) for the ideal case (left side) and
the corresponding histograms for the realistic case (right
side) are displayed.  The resulting probability-density
function (PDF) for a fixed value of frequency is shown in
logarithmic bins ($10$ per decade).

The ideal case LSP (Fig.~\ref{fig_2dhistogram}, upper panel
left side) reproduces the simulated shape of a broken
power-law with a visible deviation at high frequencies
($>0.1$~days$^{-1}$) related to the sampling frequency of
$0.297$~days$^{-1}$.  The LSPs from the uneven sampled light
curves (right side) show a considerable flattening deviating
from the original red noise behaviour.  Additionally, the
position of the break is smeared out over roughly half a
decade in frequency.  The flat (white) noise behaviour is
distorted as well, showing a decrease of the LSP for
decreasing frequencies. Besides the broad band distortion of
the LSP, features in narrow frequency bands appear as a
consequence of the sampling effects. Most notably is the
drop in the LSP at a frequency of $1$~year$^{-1}$ due to the
quasi-periodic sampling. The ideal case SF
(Fig.~\ref{fig_2dhistogram}, middle panel, left side)
follows the expected behaviour for red noise
$\mathcal{SF}\propto \tau^\alpha$ with $\alpha\approx \beta
- 1$ \citep{1992ApJ...396..469H,1993ApJ...416..485L}.  For
  values of $\tau \gtrsim 0.3 f_\mathrm{min}^{-1}$ the SF
flattens out indicating the break in the simulated
power-law.  The break in the SF is smeared out into a
gradual roll-over.  The realistic case does not deviate
systematically from the ideal case.

Finally, the result of analysing the light curve using the
MFVF shows a consistent behaviour between the ideal and
realistic simulations. In both cases, the simulated break at
$f_\mathrm{min}^{-1}=1\,000$~d translates into a bend in the
slope of the MFVF at a fragment length of $\approx
1\,000$~d.  However for the ideal case no fragments shorter
than $\approx 40\,\textrm{d}$ are found where the number of
observations is greater than $10$ due to the even sampling.
Therefore the MFVF is not defined at those time scales.

\subsection{Best-fit estimates for $f_\mathrm{min}$ and
$\beta$} \label{sub:logli} Given that the sampling of the
light-curve is generally not even, the results of the
methods described above suffer from sampling and
instrumental effects.  Therefore, it is necessary to apply a
forward folding technique to reliably reconstruct
$f_\mathrm{min}$ and $\beta$.  We have simulated $5\,000$
realistic light curves (as described in Section
\ref{sec:comparison_sim_lc}) for each parameter set on a
grid of values of $0.8 \le\beta\le 7.2$ in steps of $\Delta
\beta=0.2$ and $-4.8 \le
\log_{10}(f_\mathrm{min}/\mathrm{d}^{-1}) \le -2.2$
increasing in steps of $0.2$.  For each set of parameters we
have generated the PDFs for the LSP, SF and MFVF as
described in Section \ref{sec:comparison_sim_lc}.  For the
measured light curve a likelihood-estimator \begin{eqnarray}
\mathcal{L}(\beta,f_\mathrm{min}) &=& \sum\limits_{i=1}^{N}
\log p_i(\beta,f_\mathrm{min}), \end{eqnarray} is calculated
with $p_i(\beta,f_\mathrm{min})$ representing the
probability to measure a value of e.g. the
$\mathcal{LSP}(f_i)$ for a given set of parameters $\beta$,
$f_\mathrm{min}$ in the $i$-th bin.
$p_i(\beta,f_\mathrm{min})$ is looked-up directly in the
corresponding $\mathrm{PDF}(\beta,f_\mathrm{min})$.  The
best-fit $\mathrm{PDF}(\beta,f_\mathrm{min})$ maximises
$\mathcal{L}(\beta,f_\mathrm{min})$ and determines the
best-fit parameter set. It is noteworthy that the PDFs are
not necessarily symmetric which requires the use of a
likelihood instead of a least-squares function
\citep{2010MNRAS.404..931E}.  However, given that the
individual probabilities are not independent of each other,
the likelihood function is not well-defined.  Nevertheless,
the parameters which maximise the likelihood are good
estimators for the best-fit parameters.  It is not
straightforward to estimate the uncertainties on the
best-fit parameters given that the shape of the likelihood
function (but not the position of its maximum) are
considerably modified by the correlations between the
individual probabilities (see also
Section~\ref{sec:reliability} on tests of the method and
estimates for the uncertainties of the parameters).  Since
all simulated light curves are normalised to the same mean
flux level and variance of the measured light curve, the
methods are not sensitive to these parameters.

\section{Results} \label{sec:results} \subsection{Long-term
optical light curve of PKS~2155-304}
\label{sec:results_lightcurve}

\begin{figure*}[htb!]
\includegraphics[angle=90,width=\linewidth]{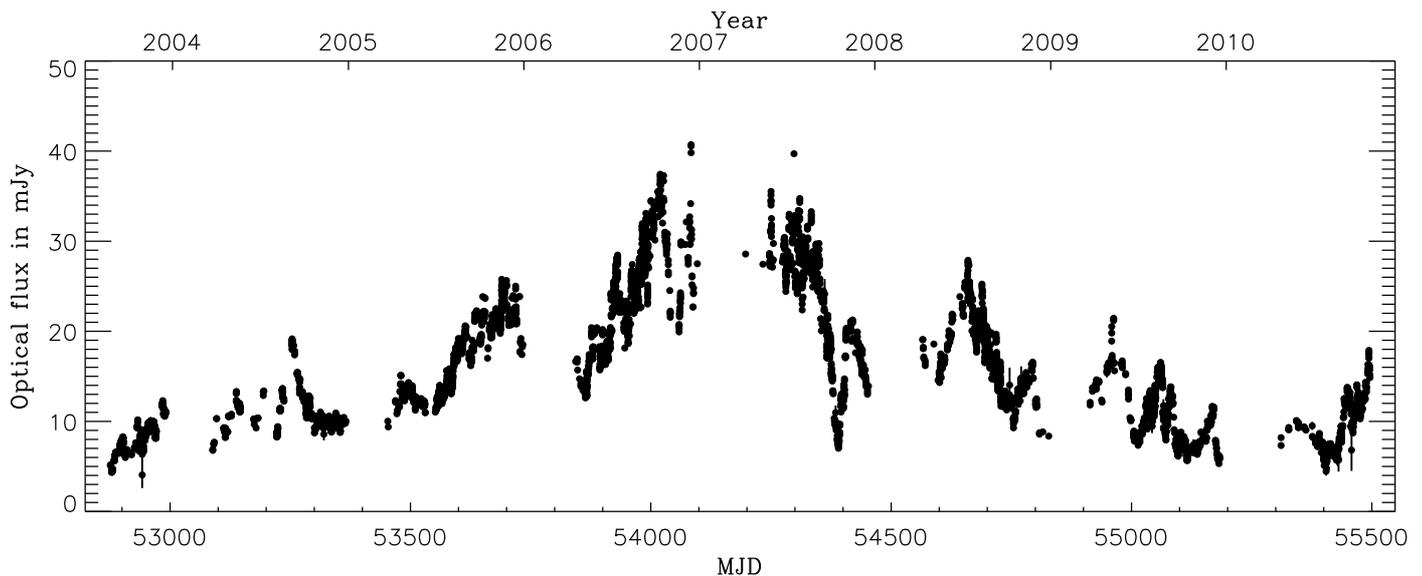}
\caption{Optical light curve (R-band) of PKS~2155-304
measured with ROTSE.  } \label{fig_rotse_lightcurve}
\end{figure*}

\begin{figure*}[htb!]
\includegraphics[angle=90,width=\linewidth]{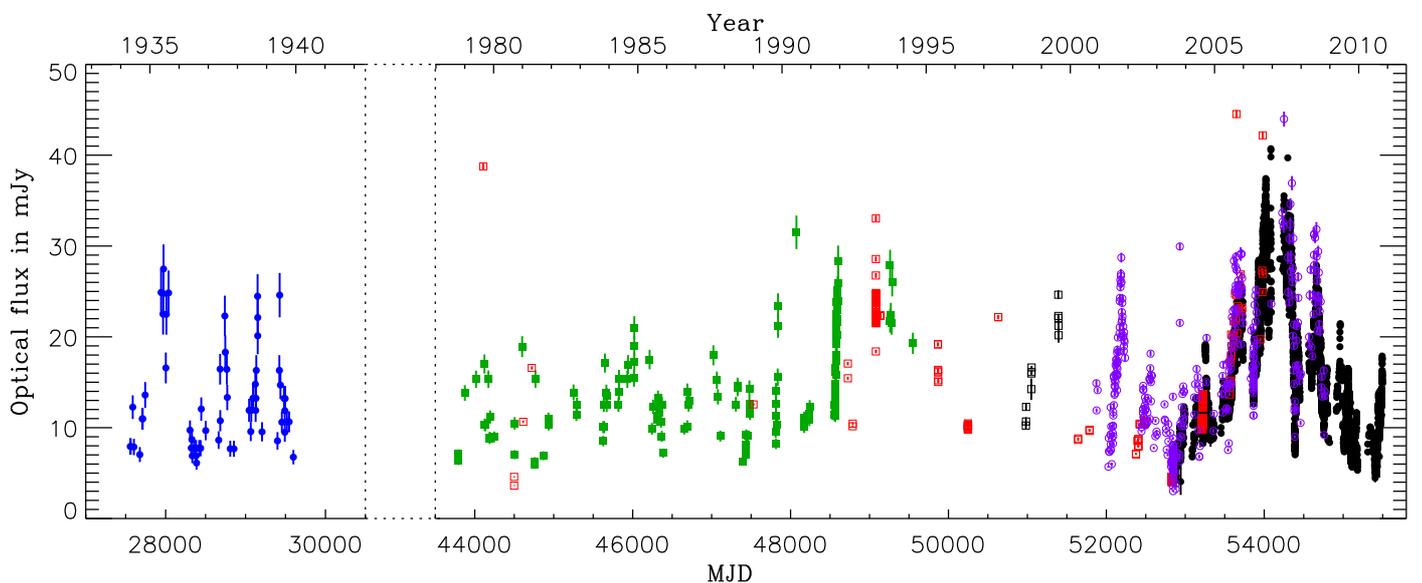}
\caption{Optical long-term light curve (R-band) of the core
region of PKS~2155-304. The light curve combines ROTSE
observations as well as other archival data (solid black
circles: ROTSE, open purple circles:
\textit{All-Sky-Automated-System}, red open squares:
\citet{2007ApJ...671...97O}, green solid squares:
\citet{1996A&AS..116..289Z}, blue solid circles:
\citet{1979ApJ...234..810G}, black open squares:
\citet{2001ApJS..132...73T}).
\label{fig_joint_longterm_kern} }\end{figure*}

The optical R-band light curve measured with ROTSE is shown
in Figure \ref{fig_rotse_lightcurve}.  The contribution of
the host galaxy ($15.1^\mathrm{mag}$) is subtracted here and
in the following. The gaps in the light curve are caused by
bad weather periods or when PKS~2155-304's position is too
close to the sun. The light curve consists of $6\,310$
individual data points.  The peak to peak amplitude observed
during this time interval varies from $5$ to $40
\,\textrm{mJy}$ indicating a relative variability of $35
\,\textrm{mJy}$ or a factor for the flux of $\approx 8$. The
variability amplitude is largest when comparing flux states
separated by long time intervals (few years) while it
decreases when comparing flux states on shorter timescales.
This behaviour is typical of red noise.

In addition to the ROTSE-III measurements archival and
contemporaneous data on PKS~2155-304, as described in
Section~\ref{sect:observations}, have been included in
Fig.~\ref{fig_joint_longterm_kern}.  Overall, the light
curve covers the time from 1934 to 2010 with an extended gap
between 1940 and 1979.  Apparently, the maximum amplitude of
variability in the long-term light curve is similar to the
observed variability in the seven years of ROTSE-III
observations.  Before analysing the light curve in detail,
this is a first indication that the red noise behaviour does
not extend towards time scales much longer than a few
decades.\\

\subsection{Results of the time series analysis}
\label{sec_application_of_methods} We have applied the three
different methods described in
Sections~\ref{sub:LSP}, \ref{sub:SF}, \ref{sub:mlvf} to analyse the
variability of the long-term light curve shown in
Fig.~\ref{fig_joint_longterm_kern}.  Based upon the
resulting LSP, SF, and MFVF, the best-fit estimators for the
power-law index $\beta$ as well as the break frequency
$f_\mathrm{min}$ have been derived individually following a
log-likelihood method described in Section~\ref{sub:logli} 

The instrumental noise for the archival data might be under-
or overestimated.  Therefore the variability on small time
scales of a simulated light curve could be distorted to an
unrealistic case.  So we use only bins at time scales
$>20$~days for the log-likelihood method to minimise
systematic errors arising from this uncertainty.

\begin{figure}[htb!]
\includegraphics[angle=90,width=\linewidth]{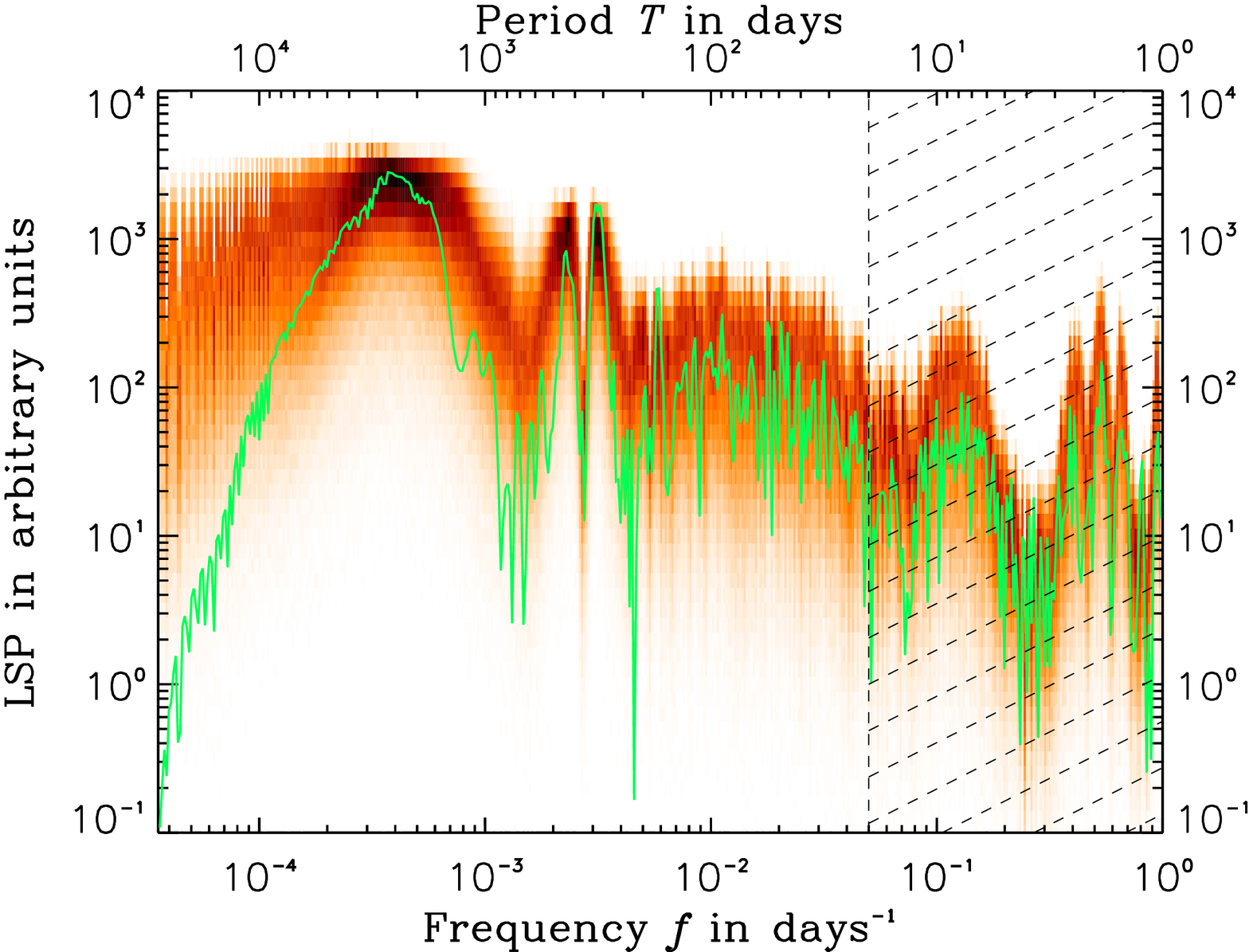}
\vskip 12pt
\includegraphics[angle=90,width=\linewidth]{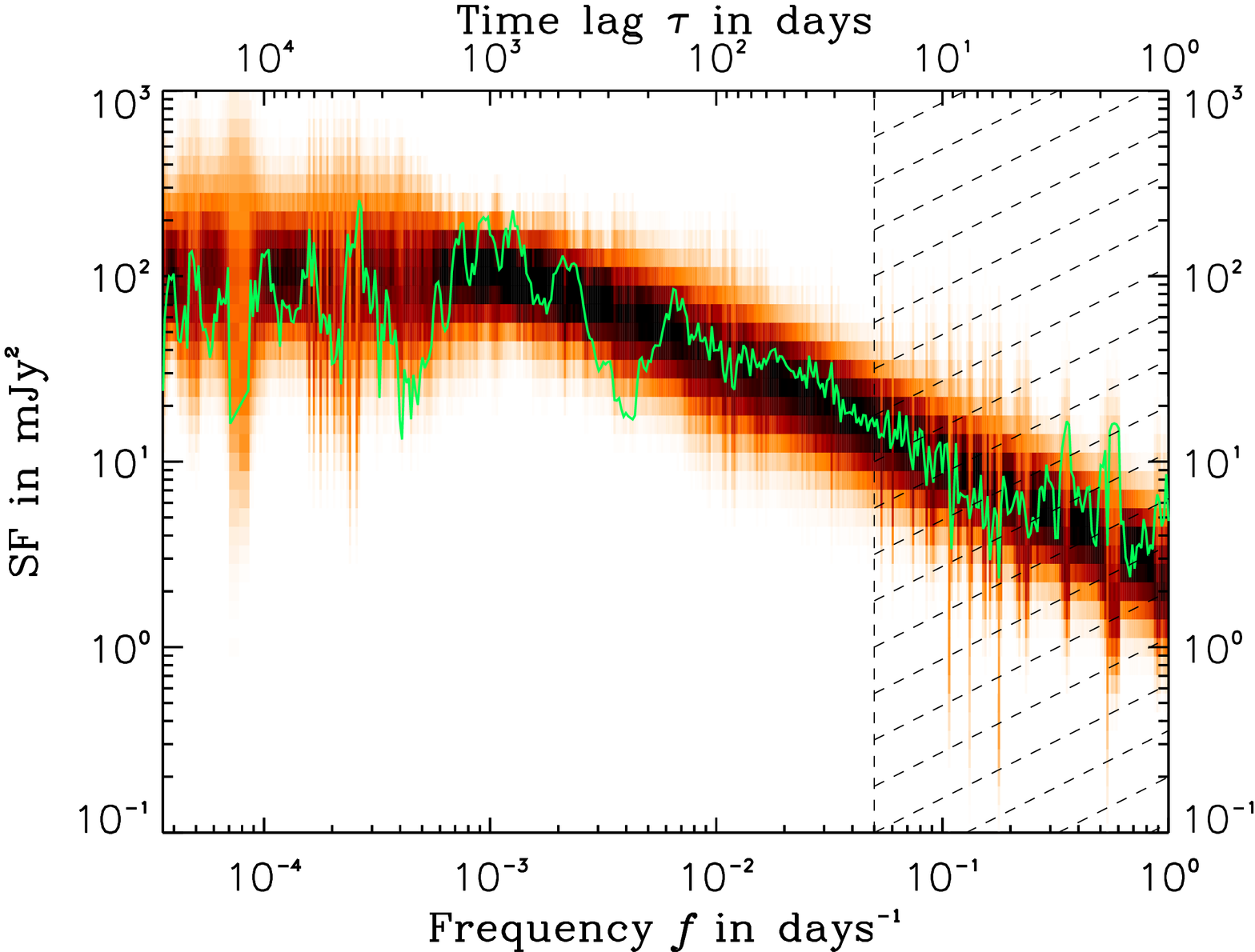}
\vskip 12pt
\includegraphics[angle=90,width=\linewidth]{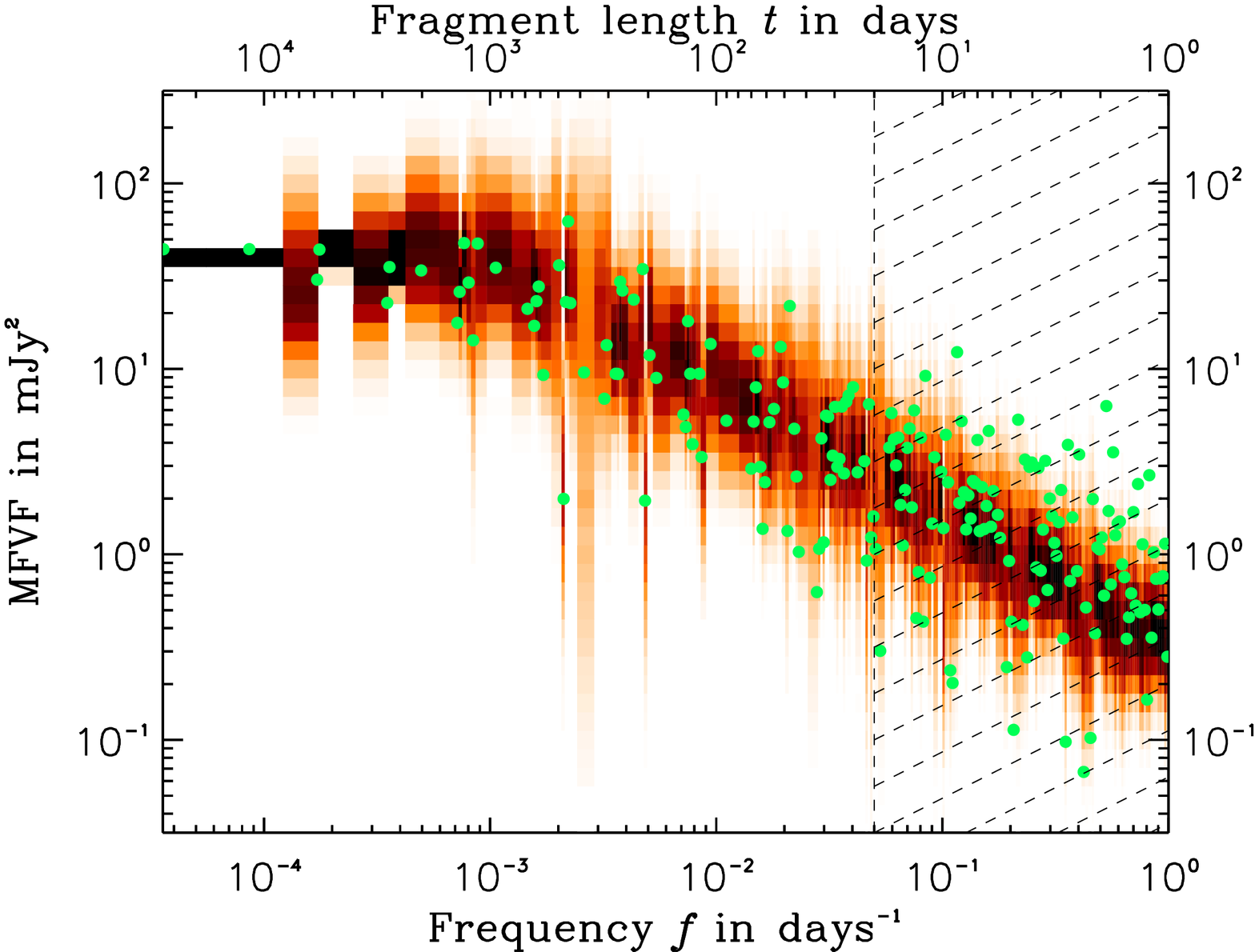}
\caption{Normalised Lomb Scargle Periodogram (LSP),
Structure Function (SF) and Multiple Fragments Variance
Function (MFVF) (from top to bottom). The green lines/points
are the measured results of the light curve of PKS~2155-304.
The histograms in colour scale represent the PDFs obtained
from simulated light curves with the best-fit parameter
sets $(\beta=5.0,\,f_\mathrm{min}=10^{-3.2}$~d$^{-1})$,
$(\beta=1.6,\,f_\mathrm{min}=10^{-3.4}$~d$^{-1})$ and
$(\beta=1.8,\,f_\mathrm{min}=10^{-3.0}$~d$^{-1})$
respectively. Only the bins at $f<0.05\,\mathrm{d}^{-1}$ are
used for the maximum-likelihood method.}
\label{fig_rotse_lsp_sf_mfvf} \end{figure}

The LSP of the long-term light curve of PKS~2155-304 is
plotted in the top panel of Fig.~\ref{fig_rotse_lsp_sf_mfvf}
together with the PDF (histogram in colour scale) of the
LSPs of the simulated light curves with the best-fitting
parameters.  The shape of the measured LSPs is in good
agreement with the PDFs derived from simulations.  As
expected, sampling effects lead to a considerable distortion
of the intrinsic power-law shape.  Additionally, the uneven
sampling produces features in the LSP which again are
reproduced in the simulated data-sets. 

In a similar fashion, the SF (middle panel in
Fig.~\ref{fig_rotse_lsp_sf_mfvf}) as well as the MFVF
(bottom panel in Fig.~\ref{fig_rotse_lsp_sf_mfvf}) are shown
together with the PDF for the best fit.  For both cases, the
simulated and measured data-sets agree well with each other.
Both, broad band shape and narrowband features from
sampling effects are well reproduced in the simulated data.  

For all methods, the resulting best-fit parameters for
$\beta$ and $f_\mathrm{min}$ with their $1\sigma$
uncertainties are listed in
Table~\ref{tab_best_fit_parameters}. The uncertainties for
the parameters are estimated as described in Section
\ref{sec:reliability}.  Graphs with the log-likelihood
functions for the parameters can be found in Appendix
\ref{appendix_likelihood}.

\begin{table} \caption{Best-fit parameters estimated by the
methods separately.} \label{tab_best_fit_parameters}
\centering \begin{tabular}{c c c} \hline \hline Method &
$\beta$ & $\log (f_\mathrm{min}/\mathrm{d}^{-1})$ \\ \hline
LSP & ${5.0}^{+1.7}_{-3.0}$ & ${-3.2}^{+0.1}_{-0.4}$ \\ SF &
${1.6}^{+0.4}_{-0.2}$ & ${-3.4}^{+0.4}_{-0.5}$  \\ MFVF &
${1.8}^{+0.1}_{-0.2}$ & ${-3.0}^{+0.3}_{-0.4}$ \\ \hline
\end{tabular} \end{table}

\subsection{Reliability of the method and estimates of the
uncertainties} \label{sec:reliability}

\begin{figure}[htb!]
\includegraphics[angle=90,width=\linewidth]{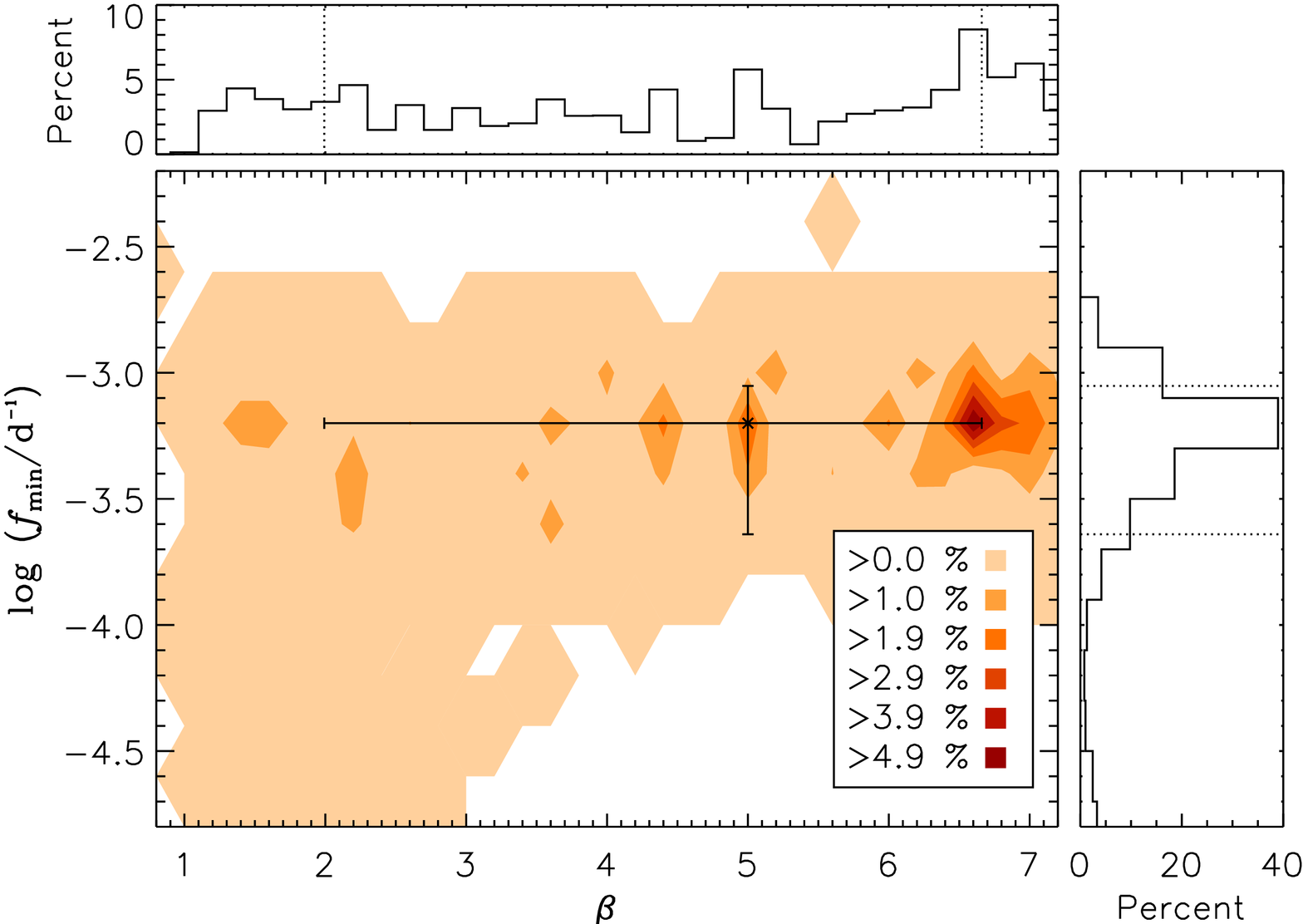}
\vskip 12pt
\includegraphics[angle=90,width=\linewidth]{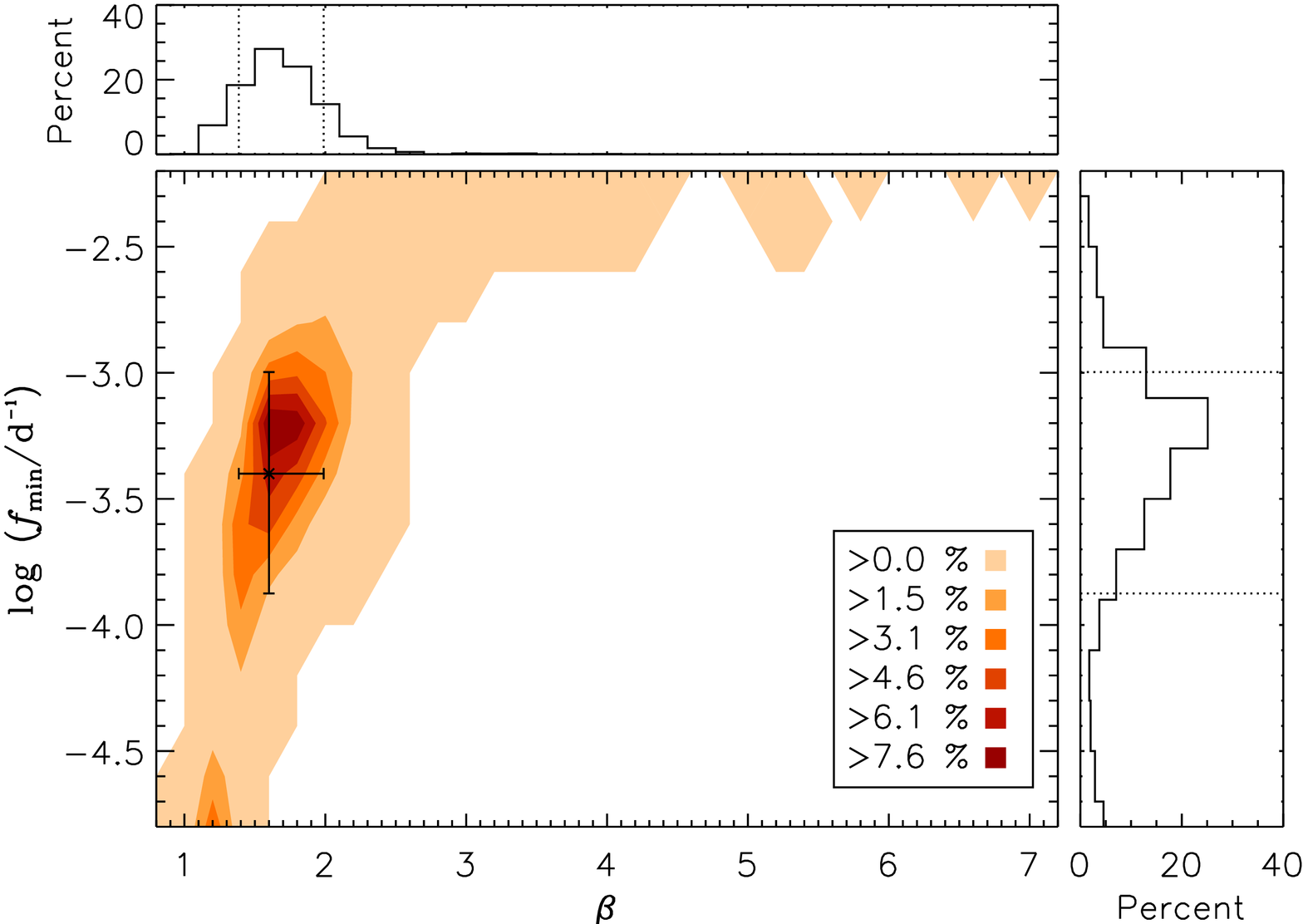}
\vskip 12pt
\includegraphics[angle=90,width=\linewidth]{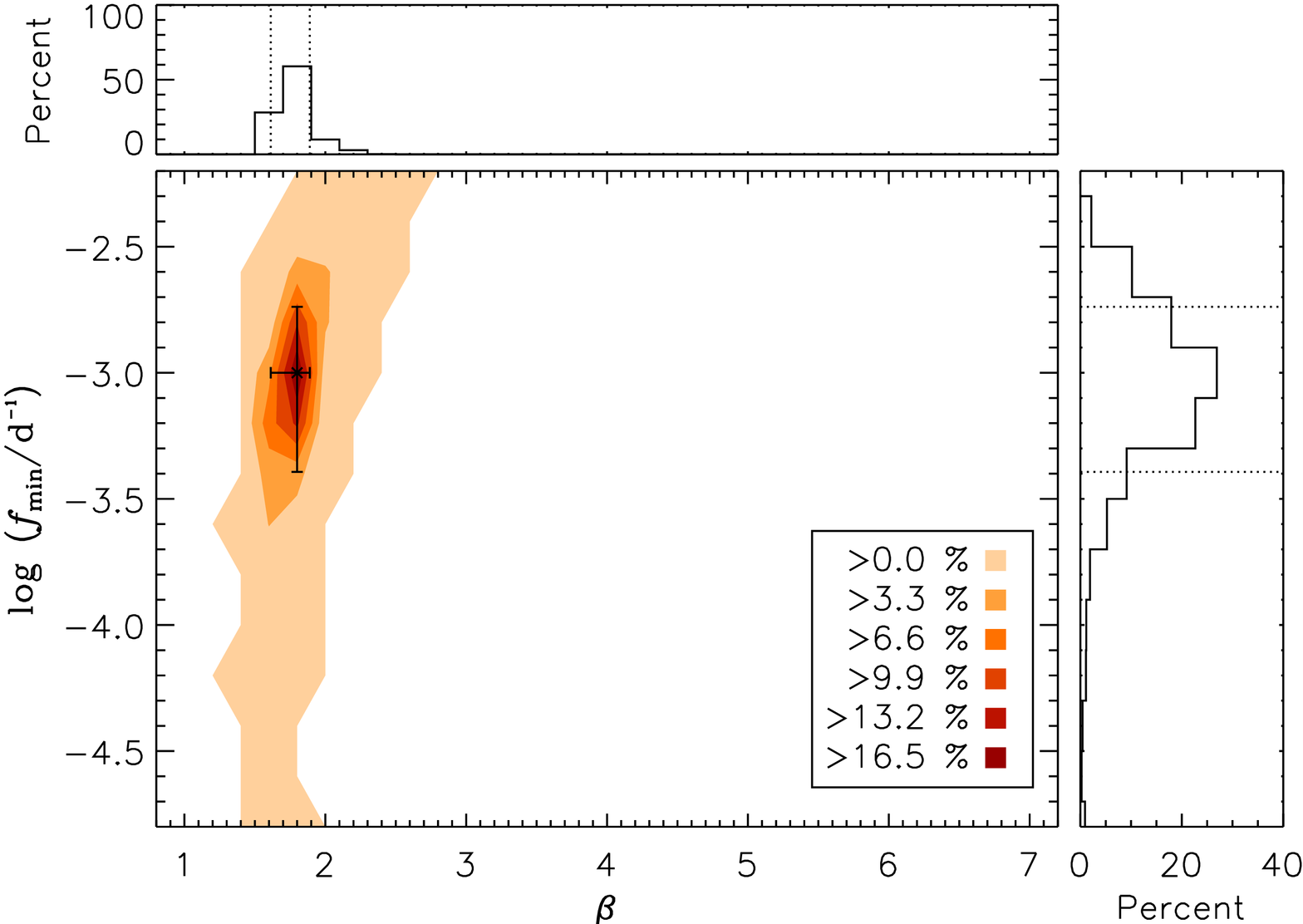}
\caption{Distribution of maximum likelihood estimates of
simulated light curves with parameters found for
PKS~2155-304. The boundaries of the coloured areas are the
contours which are quoted as fractions of the $5\,000$
simulated light curves in the legends. The sub plots show
the marginalised distributions which are calculated to
obtain the $1\,\sigma$ error bars.} \label{fig_contour}
\end{figure}

The likelihood method introduced above suffers from complex
correlations between individual probabilities. This in turn
requires to confirm that the method of maximising the
likelihood estimator does actually provide a meaningful
estimator of the underlying true parameters.  Furthermore,
it is necessary to estimate the statistical uncertainties on
the best-fit parameters without relying on the usual method
of calculating the second derivatives of the likelihood in
the maximum. \\ Therefore an empiric approach is indicated.
We have tested the reliability of the likelihood estimation
with simulated light curves created with certain (well
known) parameter sets. Again, 5\,000 realistic light curves
(as described in Section \ref{sec:comparison_sim_lc}) have
been simulated with the best-fit parameter set which has
been found with the respective method.  The light curves are
analysed with the same methods as applied to the real light
curve resulting in a distribution of best-fit parameter
sets. The resulting distributions are then used to verify
that the simulated parameters are reconstructed reliably and
the width of the distributions provide a robust estimate of
the uncertainties on the parameters.  In
Fig.~\ref{fig_contour} the two-dimensional histograms are
displayed as contours (from top to bottom LSP, SF, and
MFVF).  The cross marks the simulated (input) parameters and
the error bars indicate the 1~$\sigma$ confidence region for
each parameter of the marginalised
distributions\footnote{The 1~$\sigma$ intervals have been
calculated by integrating over one of the two parameters.
The resulting one dimensional histogram is then normalised
and the interval is truncated by taking $15.85\%$ from both
sides symmetrically such that the central interval contains
68.3~\% of the distribution} which are plotted along the
axes.\\ For the SF and MFVF, the input parameters are well
recovered with relatively small uncertainties (see
Table~\ref{tab_best_fit_parameters}) while for the LSP, the
reconstructed values are spread out over a large interval of
$\beta$ resulting in large errors.  While $f_\mathrm{min}$
is in very good agreement with the results of the SF and the
MFVF, $\beta$ differs noticeably.  However, this is in
accordance to the large uncertainties of $\beta$ for the
LSP, which is obviously affected by sampling effects. The
LSP's uncertainties for $\beta$ in the direction for higher
values is probably underestimated, because the border of the
grid is too close to the maximum of the likelihood. 

It remains to be demonstrated that the model and its
best-fit parameters are a good characterisation of the data.
Again, we use an approach relying on $5\,000$ simulated
light curves to check the consistency of data and assumed
model.  As a goodness of fit estimate the value of
$\mathcal{L}$ of the observed light curve is compared to
the distribution of $\mathcal{L}$ of the light curves
simulated with the best-fitting values of $\beta$ and
$f_\mathrm{min}$. For the three different methods a fraction
of $4~$\% (LSP), $3$~\% (SF), and $50$~\% (MFVF) of the
simulated data are fit worse by the model than the actual
data. Therefore the model is a reasonable description of the
variability of the optical flux of PKS~2155-304.

\section{Conclusion} \label{sec:summary} We have presented
an optical light-curve covering seven years of densely
sampled (6\,310 individual pointings) observations  of
PKS~2155-304 with the ROTSE-III telescopes in Namibia and
Australia.  Combined with archival data and ASAS
observations, a long-term (sparsely sampled) light-curve
going back to 1934 is analysed to characterise the
red noise behaviour under the assumption that the power
spectral density function (PSD) follows a power-law with
slope $\beta$ and a break at a minimum frequency
$f_\mathrm{min}$. At frequencies below $f_\mathrm{min}$, a
white noise ($\beta=0$) PSD is assumed.  For the analysis
only time scales $>20$~days have been considered to avoid
systematic errors due to uncertain estimates of the
instrumental noise for the simulated light curves.  The three
different methods for analysing the long-term variability of
PKS~2155-304 produce consistent results for simulated light
curves with $\beta\approx 1.8$ and $f_\mathrm{min}\approx
1/1\,000~$d$^{-1}$. The most precise result (discussed in
the following) is given by the newly introduced Multiple
Fragments Variance Function (MFVF), which is a
straightforward method in the time domain to reveal
information about the mean variability amplitude depending
on different time scales. The result obtained with the
first-order Structure Function (SF) is in very good
agreement with the result of the MFVF.  The slope of the
Lomb-Scargle Periodogram (LSP) is affected strongly by
sampling effects, so the uncertainties of the best-fit
parameter $\beta$ are considerably larger for this method
than for the MFVF and SF.

The power-law index of the PSD $\beta=1.8^{+0.1}_{-0.2}$
is harder than the value obtained by
\citet{1997A&A...327..539P} at timescales smaller than 10~d
($\beta=2.4^{+0.3}_{-0.2}$). This may be an indication for
an additional break or curvature in the PSD on short
timescales. The power-law index of the red noise observed
at other energy scales are similar.  This hints at a common
origin of the observed emission at the base of the jet.

Our analysis confirms the presence of red noise up to
timescales of a few years without significant indications
for any additional features as e.g.  quasi-periodic
oscillations.  The existence of a break in the PSD at
$f_\mathrm{min}=1/1\,000\,\mathrm{d}^{-1}$ is found to be
significant at the level of $\approx 6~\sigma$ judging from
the error on $f_\mathrm{min}$ estimated for the MFVF method.
This break corresponds to a time scale of $\tau_\mathrm{max}
\approx 2.7^{+2.2}_{-1.6}\,\textrm{yrs}$ after which the red
noise behaviour disappears.  In the context of SOC this time
scale is linked to the ``memory'' of the system. At time
scales longer than $\tau_\mathrm{max}$ the properties of the
emission process de-correlate from its past.  Due to causal
reasons, the  spatial dimension of the emission region is
constrained to be
$r<c\tau_\mathrm{max}=0.84^{+0.69}_{-0.50}~\,$pc.

The observed red-noise from PKS~2155-304 extends over five
orders of magnitude in the time domain and over twelve
orders of magnitude in the photon energy with apparently
similar power-law slope. This implies that there is probably
a link connecting particle acceleration processes on spatial
scales ranging from ${\cal O}(100\,\mathrm{AU})$ to ${\cal
O}(1\,\mathrm{pc})$.

\begin{acknowledgements} We thank the entire ROTSE team
including the local support in Namibia (Toni Hanke) and
Australia (Andre Philips and Michael Ashley) for the
excellent support and for the opportunity to use observation
time on the ROTSE-IIIa and c telescopes.  M.A. K.
acknowledges the financial support of the BMBF under the
contract number 05A08GU1 and the support of the Hamburg
cluster of excellence Connecting Particles with the Cosmos.
We thank Martin Raue for useful suggestions to improve the
manuscript. This research has made use of NASA’s
Astrophysics Data System and NASA/IPAC Extragalactic
Database (NED) which is operated by the Jet Propulsion
Laboratory, California Institute of Technology, under
contract with the National Aeronautics and Space
Administration. Thanks to Marcus Hauser from the Automatic
Telescope for Optical Monitoring (ATOM) for assisting the
cross calibration of the light curves.
\end{acknowledgements}

\bibliography{referenzen} \bibliographystyle{aa}

\begin{thebibliography}{45}
\expandafter\ifx\csname natexlab\endcsname\relax\def\natexlab#1{#1}\fi

\bibitem[{{Aharonian} {et~al.}(2009){Aharonian}, {Akhperjanian}, {Anton},
  {Barres de Almeida}, {Bazer-Bachi}, {Becherini}, {Behera}, {Bernl{\"o}hr},
  {Boisson}, {Bochow}, \& et~al.}]{2009ApJ...696L.150A}
{Aharonian}, F., {Akhperjanian}, A.~G., {Anton}, G., {et~al.} 2009, \apjl, 696,
  L150

\bibitem[{{Aharonian} {et~al.}(2007){Aharonian}, {Akhperjanian}, {Bazer-Bachi},
  {Behera}, {Beilicke}, {Benbow}, {Berge}, {Bernl{\"o}hr}, {Boisson}, {Bolz},
  {Borrel}, {Boutelier}, {Braun}, {Brion}, {Brown}, {B{\"u}hler},
  {B{\"u}sching}, {Bulik}, {Carrigan}, {Chadwick}, {Clapson}, {Chounet},
  {Coignet}, {Cornils}, {Costamante}, {Degrange}, {Dickinson},
  {Djannati-Ata{\"i}}, {Domainko}, {Drury}, {Dubus}, {Dyks}, {Egberts},
  {Emmanoulopoulos}, {Espigat}, {Farnier}, {Feinstein}, {Fiasson},
  {F{\"o}rster}, {Fontaine}, {Funk}, {Funk}, {F{\"u}{\ss}ling}, {Gallant},
  {Giebels}, {Glicenstein}, {Gl{\"u}ck}, {Goret}, {Hadjichristidis}, {Hauser},
  {Hauser}, {Heinzelmann}, {Henri}, {Hermann}, {Hinton}, {Hoffmann}, {Hofmann},
  {Holleran}, {Hoppe}, {Horns}, {Jacholkowska}, {de Jager}, {Kendziorra},
  {Kerschhaggl}, {Kh{\'e}lifi}, {Komin}, {Kosack}, {Lamanna}, {Latham}, {Le
  Gallou}, {Lemi{\`e}re}, {Lemoine-Goumard}, {Lenain}, {Lohse}, {Martin},
  {Martineau-Huynh}, {Marcowith}, {Masterson}, {Maurin}, {McComb}, {Moderski},
  {Moulin}, {de Naurois}, {Nedbal}, {Nolan}, {Olive}, {Orford}, {Osborne},
  {Ostrowski}, {Panter}, {Pedaletti}, {Pelletier}, {Petrucci}, {Pita},
  {P{\"u}hlhofer}, {Punch}, {Ranchon}, {Raubenheimer}, {Raue}, {Rayner},
  {Renaud}, {Ripken}, {Rob}, {Rolland}, {Rosier-Lees}, {Rowell}, {Rudak},
  {Ruppel}, {Sahakian}, {Santangelo}, {Saug{\'e}}, {Schlenker}, {Schlickeiser},
  {Schr{\"o}der}, {Schwanke}, {Schwarzburg}, {Schwemmer}, {Shalchi}, {Sol},
  {Spangler}, {Stawarz}, {Steenkamp}, {Stegmann}, {Superina}, {Tam},
  {Tavernet}, {Terrier}, {van Eldik}, {Vasileiadis}, {Venter}, {Vialle},
  {Vincent}, {Vivier}, {V{\"o}lk}, {Volpe}, {Wagner}, {Ward}, \&
  {Zdziarski}}]{2007ApJ...664L..71A}
{Aharonian}, F., {Akhperjanian}, A.~G., {Bazer-Bachi}, A.~R., {et~al.} 2007,
  \apjl, 664, L71

\bibitem[{{Aharonian} {et~al.}(2005){Aharonian}, {Akhperjanian}, {Bazer-Bachi},
  {Beilicke}, {Benbow}, {Berge}, {Bernl{\"o}hr}, {Boisson}, {Bolz}, {Borrel},
  {Braun}, {Breitling}, {Brown}, {Chadwick}, {Chounet}, {Cornils},
  {Costamante}, {Degrange}, {Dickinson}, {Djannati-Ata{\"i}}, {O'C.~Drury},
  {Dubus}, {Emmanoulopoulos}, {Espigat}, {Feinstein}, {Fontaine}, {Fuchs},
  {Funk}, {Gallant}, {Giebels}, {Gillessen}, {Glicenstein}, {Goret},
  {Hadjichristidis}, {Hauser}, {Heinzelmann}, {Henri}, {Hermann}, {Hinton},
  {Hofmann}, {Holleran}, {Horns}, {Jacholkowska}, {de Jager}, {Kh{\'e}lifi},
  {Komin}, {Konopelko}, {Latham}, {Le Gallou}, {Lemi{\`e}re},
  {Lemoine-Goumard}, {Leroy}, {Lohse}, {Martin}, {Martineau-Huynh},
  {Marcowith}, {Masterson}, {McComb}, {de Naurois}, {Nolan}, {Noutsos},
  {Orford}, {Osborne}, {Ouchrif}, {Panter}, {Pelletier}, {Pita},
  {P{\"u}hlhofer}, {Punch}, {Raubenheimer}, {Raue}, {Raux}, {Rayner}, {Reimer},
  {Reimer}, {Ripken}, {Rob}, {Rolland}, {Rowell}, {Sahakian}, {Saug{\'e}},
  {Schlenker}, {Schlickeiser}, {Schuster}, {Schwanke}, {Siewert}, {Sol},
  {Spangler}, {Steenkamp}, {Stegmann}, {Tavernet}, {Terrier}, {Th{\'e}oret},
  {Tluczykont}, {Vasileiadis}, {Venter}, {Vincent}, {V{\"o}lk}, \&
  {Wagner}}]{2005A&A...442..895A}
{Aharonian}, F., {Akhperjanian}, A.~G., {Bazer-Bachi}, A.~R., {et~al.} 2005,
  \aap, 442, 895

\bibitem[{{Akerlof} {et~al.}(2003){Akerlof}, {Kehoe}, {McKay}, {Rykoff},
  {Smith}, {Casperson}, {McGowan}, {Vestrand}, {Wozniak}, {Wren}, {Ashley},
  {Phillips}, {Marshall}, {Epps}, \& {Schier}}]{2003PASP..115..132A}
{Akerlof}, C.~W., {Kehoe}, R.~L., {McKay}, T.~A., {et~al.} 2003, \pasp, 115,
  132

\bibitem[{{Bak} {et~al.}(1987){Bak}, {Tang}, \&
  {Wiesenfeld}}]{1987PhRvL..59..381B}
{Bak}, P., {Tang}, C., \& {Wiesenfeld}, K. 1987, Physical Review Letters, 59,
  381

\bibitem[{{Bertin}(2002)}]{sextractor_anleitung}
{Bertin}, E. 2002, SExtractor User's guide, Institut d'Astrophysique \&
  Observatoire de Paris

\bibitem[{{Brindle} {et~al.}(1986){Brindle}, {Hough}, {Bailey}, {Axon}, \&
  {Hyland}}]{1986MNRAS.221..739B}
{Brindle}, C., {Hough}, J.~H., {Bailey}, J.~A., {Axon}, D.~J., \& {Hyland},
  A.~R. 1986, \mnras, 221, 739

\bibitem[{{Campbell}(2004)}]{2004PhDT........28C}
{Campbell}, A.~M. 2004, PhD thesis, Georgia State University, United States --
  Georgia

\bibitem[{{Carini} \& {Miller}(1992)}]{1992ApJ...385..146C}
{Carini}, M.~T. \& {Miller}, H.~R. 1992, \apj, 385, 146

\bibitem[{{Courvoisier} {et~al.}(1995){Courvoisier}, {Blecha}, {Bouchet},
  {Bratschi}, {Carini}, {Donahue}, {Edelson}, {Feigelson}, {Filippenko},
  {Glass}, {Heidt}, {Kollgaard}, {Matheson}, {Miller}, {Noble}, {Sekiguchi},
  {Smith}, {Urry}, \& {Wagner}}]{1995ApJ...438..108C}
{Courvoisier}, T., {Blecha}, A., {Bouchet}, P., {et~al.} 1995, \apj, 438, 108

\bibitem[{{Dendy} {et~al.}(1999){Dendy}, {Helander}, \&
  {Tagger}}]{1999PhST...82..133D}
{Dendy}, R.~O., {Helander}, P., \& {Tagger}, M. 1999, Physica Scripta Volume T,
  82, 133

\bibitem[{{Emmanoulopoulos} {et~al.}(2010){Emmanoulopoulos}, {McHardy}, \&
  {Uttley}}]{2010MNRAS.404..931E}
{Emmanoulopoulos}, D., {McHardy}, I.~M., \& {Uttley}, P. 2010, \mnras, 404, 931

\bibitem[{{Falomo} {et~al.}(1993){Falomo}, {Pesce}, \&
  {Treves}}]{1993ApJ...411L..63F}
{Falomo}, R., {Pesce}, J.~E., \& {Treves}, A. 1993, \apjl, 411, L63

\bibitem[{{Gaur} {et~al.}(2010){Gaur}, {Gupta}, {Lachowicz}, \&
  {Wiita}}]{2010ApJ...718..279G}
{Gaur}, H., {Gupta}, A.~C., {Lachowicz}, P., \& {Wiita}, P.~J. 2010, \apj, 718,
  279

\bibitem[{{Griffiths} {et~al.}(1979){Griffiths}, {Briel}, {Chaisson}, \&
  {Tapia}}]{1979ApJ...234..810G}
{Griffiths}, R.~E., {Briel}, U., {Chaisson}, L., \& {Tapia}, S. 1979, \apj,
  234, 810

\bibitem[{{Hamuy} \& {Maza}(1987)}]{1987A&AS...68..383H}
{Hamuy}, M. \& {Maza}, J. 1987, \aaps, 68, 383

\bibitem[{{Heidt} {et~al.}(1997){Heidt}, {Wagner}, \&
  {Wilhelm-Erkens}}]{1997A&A...325...27H}
{Heidt}, J., {Wagner}, S.~J., \& {Wilhelm-Erkens}, U. 1997, \aap, 325, 27

\bibitem[{{Hughes} {et~al.}(1992){Hughes}, {Aller}, \&
  {Aller}}]{1992ApJ...396..469H}
{Hughes}, P.~A., {Aller}, H.~D., \& {Aller}, M.~F. 1992, \apj, 396, 469

\bibitem[{{Jannuzi} {et~al.}(1993){Jannuzi}, {Smith}, \&
  {Elston}}]{1993ApJS...85..265J}
{Jannuzi}, B.~T., {Smith}, P.~S., \& {Elston}, R. 1993, \apjs, 85, 265

\bibitem[{{Kawaguchi} \& {Mineshige}(1999)}]{1999IAUS..194..356K}
{Kawaguchi}, T. \& {Mineshige}, S. 1999, in IAU Symposium, Vol. 194, Activity
  in Galaxies and Related Phenomena, ed. Y.~{Terzian}, E.~{Khachikian}, \&
  D.~{Weedman}, 356 ff.

\bibitem[{{Kotilainen} {et~al.}(1998){Kotilainen}, {Falomo}, \&
  {Scarpa}}]{1998A&A...336..479K}
{Kotilainen}, J.~K., {Falomo}, R., \& {Scarpa}, R. 1998, \aap, 336, 479

\bibitem[{{Lainela} \& {Valtaoja}(1993)}]{1993ApJ...416..485L}
{Lainela}, M. \& {Valtaoja}, E. 1993, \apj, 416, 485 ff.

\bibitem[{{Lomb}(1976)}]{1976Ap&SS..39..447L}
{Lomb}, N.~R. 1976, \apss, 39, 447

\bibitem[{{Malkov} {et~al.}(2000){Malkov}, {Diamond}, \&
  {V{\"o}lk}}]{2000ApJ...533L.171M}
{Malkov}, M.~A., {Diamond}, P.~H., \& {V{\"o}lk}, H.~J. 2000, \apjl, 533, L171

\bibitem[{{Mead} {et~al.}(1990){Mead}, {Ballard}, {Brand}, {Hough}, {Brindle},
  \& {Bailey}}]{1990A&AS...83..183M}
{Mead}, A.~R.~G., {Ballard}, K.~R., {Brand}, P.~W.~J.~L., {et~al.} 1990, \aaps,
  83, 183

\bibitem[{{Miller} \& {McAlister}(1983)}]{1983ApJ...272...26M}
{Miller}, H.~R. \& {McAlister}, H.~A. 1983, \apj, 272, 26

\bibitem[{{Osterman} {et~al.}(2007){Osterman}, {Miller}, {Marshall}, {Ryle},
  {Aller}, {Aller}, \& {McFarland}}]{2007ApJ...671...97O}
{Osterman}, M.~A., {Miller}, H.~R., {Marshall}, K., {et~al.} 2007, \apj, 671,
  97

\bibitem[{{Paltani} {et~al.}(1997){Paltani}, {Courvoisier}, {Blecha}, \&
  {Bratschi}}]{1997A&A...327..539P}
{Paltani}, S., {Courvoisier}, T.~J.-L., {Blecha}, A., \& {Bratschi}, P. 1997,
  \aap, 327, 539

\bibitem[{{Pesce} {et~al.}(1997){Pesce}, {Urry}, {Maraschi}, {Treves},
  {Grandi}, {Kollgaard}, {Pian}, {Smith}, {Aller}, {Aller}, {Barth}, {Buckley},
  {Covino}, {Filippenko}, {Hooper}, {Joner}, {Kedziora-Chudczer}, {Kilkenny},
  {Knee}, {Kunkel}, {Layden}, {Magalhaes}, {Marang}, {Margoniner}, {Palma},
  {Pereyra}, {Rodrigues}, {Schutte}, {Sitko}, {Tornikoski}, {van der Walt},
  {van Wyk}, {Whitelock}, \& {Wolk}}]{1997ApJ...486..770P}
{Pesce}, J.~E., {Urry}, C.~M., {Maraschi}, L., {et~al.} 1997, \apj, 486, 770
  ff.

\bibitem[{{Pica} {et~al.}(1988){Pica}, {Smith}, {Webb}, {Leacock}, {Clements},
  \& {Gombola}}]{1988AJ.....96.1215P}
{Pica}, A.~J., {Smith}, A.~G., {Webb}, J.~R., {et~al.} 1988, \aj, 96, 1215

\bibitem[{{Pojmanski}(2002)}]{2002AcA....52..397P}
{Pojmanski}, G. 2002, Acta Astronomica, 52, 397

\bibitem[{{Press} \& {Rybicki}(1989)}]{1989ApJ...338..277P}
{Press}, W.~H. \& {Rybicki}, G.~B. 1989, \apj, 338, 277

\bibitem[{{Rieger}(2004)}]{2004ApJ...615L...5R}
{Rieger}, F.~M. 2004, \apjl, 615, L5

\bibitem[{{Rykoff} \& {Smith}(2003)}]{rotse_anleitung}
{Rykoff}, E. \& {Smith}, D. 2003, Components and Operation of the ROTSE-III
  Telescope System, University of Michigan \\
  \texttt{http://www.rotse.net/equipment/docs/rotsedocs.pdf}

\bibitem[{{Scargle}(1982)}]{1982ApJ...263..835S}
{Scargle}, J.~D. 1982, \apj, 263, 835

\bibitem[{{Simonetti} {et~al.}(1985){Simonetti}, {Cordes}, \&
  {Heeschen}}]{1985ApJ...296...46S}
{Simonetti}, J.~H., {Cordes}, J.~M., \& {Heeschen}, D.~S. 1985, \apj, 296, 46

\bibitem[{{Sivron}(1998)}]{1998ApJ...503L..57S}
{Sivron}, R. 1998, \apjl, 503, L57+

\bibitem[{{Smith} {et~al.}(1992){Smith}, {Hall}, {Allen}, \&
  {Sitko}}]{1992ApJ...400..115S}
{Smith}, P.~S., {Hall}, P.~B., {Allen}, R.~G., \& {Sitko}, M.~L. 1992, \apj,
  400, 115

\bibitem[{{Smith} \& {Sitko}(1991)}]{1991ApJ...383..580S}
{Smith}, P.~S. \& {Sitko}, M.~L. 1991, \apj, 383, 580

\bibitem[{{Timmer} \& {Koenig}(1995)}]{1995A&A...300..707T}
{Timmer}, J. \& {Koenig}, M. 1995, \aap, 300, 707 ff.

\bibitem[{{Tommasi} {et~al.}(2001){Tommasi}, {D{\'{\i}}az}, {Palazzi}, {Pian},
  {Poretti}, {Scaltriti}, \& {Treves}}]{2001ApJS..132...73T}
{Tommasi}, L., {D{\'{\i}}az}, R., {Palazzi}, E., {et~al.} 2001, \apjs, 132, 73

\bibitem[{{Treves} {et~al.}(1989){Treves}, {Morini}, {Chiappetti}, {Fabian},
  {Falomo}, {Maccagni}, {Maraschi}, {Tanzi}, \&
  {Tagliaferri}}]{1989ApJ...341..733T}
{Treves}, A., {Morini}, M., {Chiappetti}, L., {et~al.} 1989, \apj, 341, 733

\bibitem[{{Ulrich} {et~al.}(1997){Ulrich}, {Maraschi}, \&
  {Urry}}]{1997ARA&A..35..445U}
{Ulrich}, M.-H., {Maraschi}, L., \& {Urry}, C.~M. 1997, \araa, 35, 445

\bibitem[{{Urry} {et~al.}(1993){Urry}, {Maraschi}, {Edelson}, {Koratkar},
  {Krolik}, {Madejski}, {Pian}, {Pike}, {Reichert}, {Treves}, {Wamsteker},
  {Bohlin}, {Bregman}, {Brinkmann}, {Chiappetti}, {Courvoisier}, {Filippenko},
  {Fink}, {George}, {Kondo}, {Martin}, {Miller}, {O'Brien}, {Shull}, {Sitko},
  {Szymkowiak}, {Tagliaferri}, {Wagner}, \& {Warwick}}]{1993ApJ...411..614U}
{Urry}, C.~M., {Maraschi}, L., {Edelson}, R., {et~al.} 1993, \apj, 411, 614

\bibitem[{{Zhang} \& {Xie}(1996)}]{1996A&AS..116..289Z}
{Zhang}, Y.~H. \& {Xie}, G.~Z. 1996, \aaps, 116, 289

\end{thebibliography}
\FloatBarrier

\appendix \section{Standard methods of time series analysis}
\subsection{Lomb-Scargle Periodogram (LSP)} \label{sub:LSP}
The normalised LSP
\citep{1976Ap&SS..39..447L,1982ApJ...263..835S} is a method
to estimate the true underlying PSD of the variability of
unevenly sampled time series data.  For a given set of
values $a_i:=a(t_i)$ measured at the times $t_i$ the LSP is
defined as

\begin{eqnarray} \mathcal{LSP}(\omega) = \frac{1}{2
\sigma^2} \left\{ \frac{ \left[ \sum_j \left( a_j - \bar
a\right) \, \cos \left( \omega \left( t_j - \epsilon
\right)\right) \right] ^2 }{ \sum_j \cos^2 \left(
\omega\left( t_j - \epsilon \right)\right) }\right.
\nonumber \\ + \left.\frac{ \left[ \sum_j \left( a_j - \bar
a\right) \, \sin \left( \omega \left( t_j - \epsilon
\right)\right) \right] ^2 }{ \sum_j \sin^2 \left(
\omega\left( t_j - \epsilon \right)\right) } \right\},
\end{eqnarray}

with $\omega=2\pi f$, $\sigma^2$ being  the variance and
$\bar a$ the mean value of the time-series $a_i$.  The time
delay \begin{eqnarray} \epsilon&=&\frac{1}{2\omega}
\tan^{-1} \frac{\sum_j \sin \left(2\omega t_j\right)}{\sum_j
\cos \left(2\omega t_j\right)} \end{eqnarray} is required to
achieve that $\mathcal{LSP}(\omega)$ is invariant under a
change of the phase \citep{1989ApJ...338..277P}. In order to
weigh widely different frequency scales equally, we
calculate the LSP for values of $f$ with a constant interval
of 0.01 decades on a logarithmic scale.

\subsection{Structure Function (SF)} \label{sub:SF} The
first order SF \citep{1985ApJ...296...46S} for a time-lag
$\tau$ between any pair of observations is defined as
\begin{eqnarray}\label{eq_sf} \mathcal{SF} ( \tau)=
\frac{1}{n} \sum_{t_i} \left[ a \left( t_i+\tau \right) - a
\left( t_i \right) \right]^2, \end{eqnarray} with $n$ being
the number of summands.  The structure function can only be
computed for discrete values of $\tau$ that exist in the
time series: \begin{eqnarray} \tau \in \left\{ \left|
t_i-t_j \right| : t_i, t_j \in T \right\} \; ,
\end{eqnarray} where $T$ is the set of times where
measurements have been carried out.  Commonly, the values of
the structure function are combined and averaged in
intervals of $\tau$  to smooth fluctuations on time
scales shorter than the width of the bins.  In order to
weigh widely different time scales equally, we choose a
constant bin width of $0.01$ decades on a logarithmic scale.

\section{Log likelihood} \label{appendix_likelihood}

\begin{figure}[htb!]
\includegraphics[width=\linewidth]{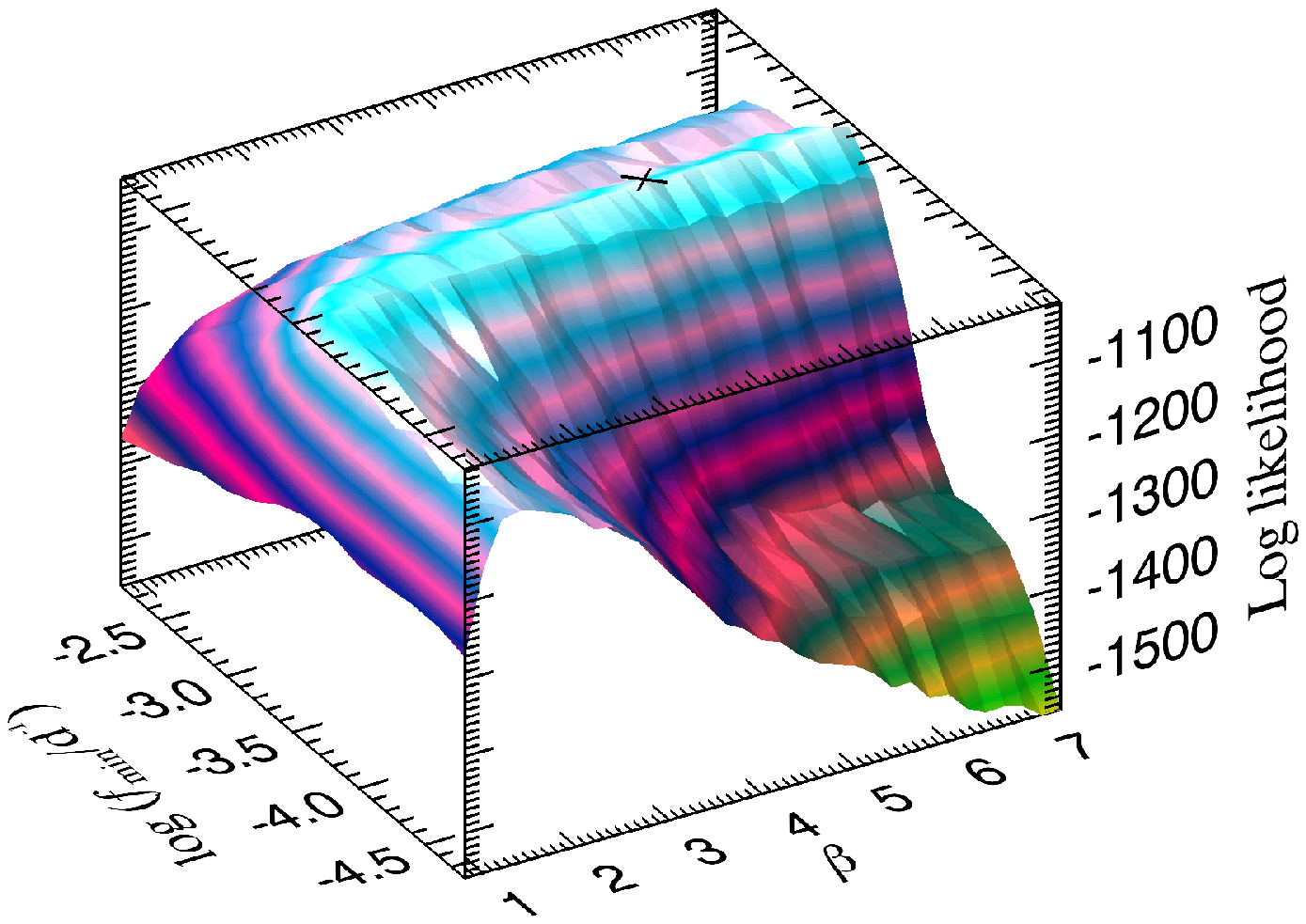}
\includegraphics[width=\linewidth]{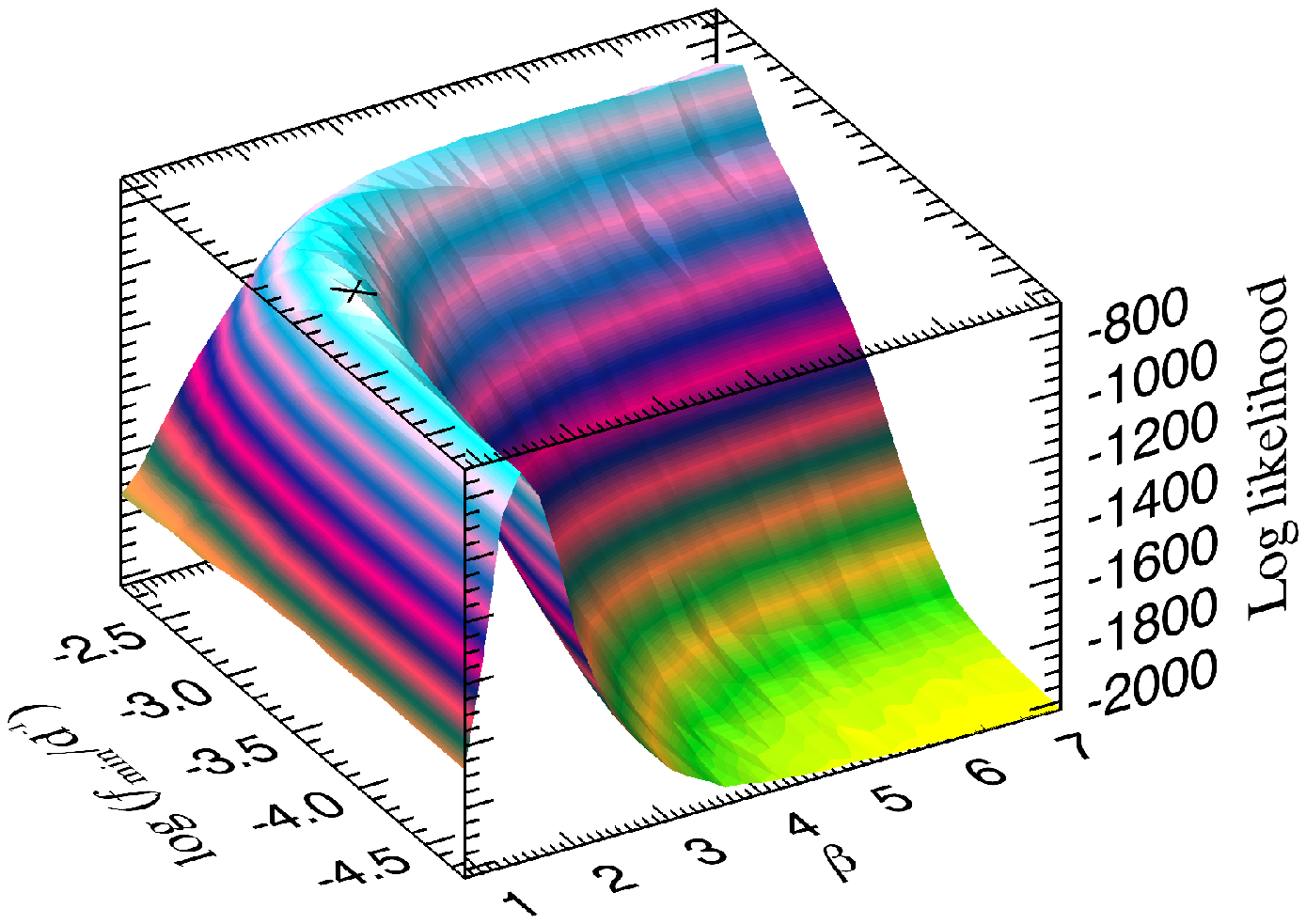}
\includegraphics[width=\linewidth]{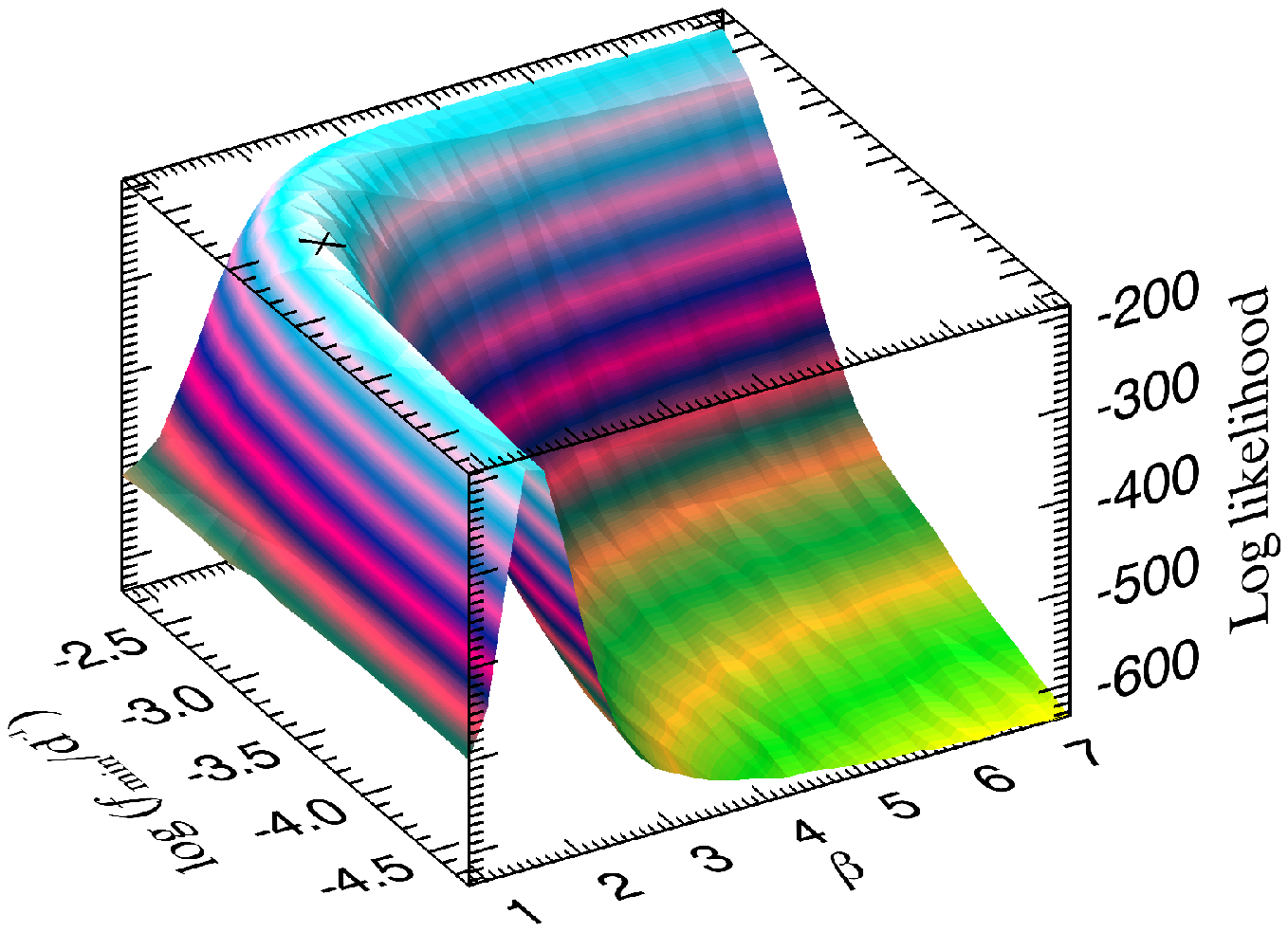}
\caption{Logarithm of the likelihood for the real light
curve being realised by red noise with a certain $\beta$ and
$f_\mathrm{min}$. Top for the LSP, middle for the SF and
bottom for the MFVF.  The peaks located in the white shaded
areas are marked with a black cross.}
\label{fig_loglikelihood} \end{figure}

The log-likelihood values for the grid of the parameters are
shown in the panels (from top to bottom LSP, SF, and MFVF)
of Fig.~\ref{fig_loglikelihood}.  The log-likelihood
functions $\mathcal{L}$ of the SF and MFVF have a similar
shape with a ridge-like region with a bend. For the SF and
the MFVF the maximum is located in the region where the
ridge bends. For the LSP, the maximum of the likelihood is
located on a ridge which is oriented along the direction of
increasing $\beta$, which obviously does not constrain the
best-fit value of $\beta$ very well.  \end{document}